\documentclass[12pt]{article}

\usepackage{setspace}
\usepackage{graphicx}
\usepackage[margin=1.9cm]{geometry}
\usepackage{lscape}
\usepackage{xcolor}
\usepackage{amsmath,amssymb}
\usepackage{hyperref} 
\usepackage{gensymb}
\usepackage[load=accepted]{siunitx}
\usepackage{lineno}
\usepackage{flafter}
\usepackage{hyperref}

\newcommand{\feii}{\mbox{Fe\,{\sc ii}}}
\newcommand{\znii}{\mbox{Zn\,{\sc ii}}}
\newcommand{\crii}{\mbox{Cr\,{\sc ii}}}
\newcommand{\tiii}{\mbox{Ti\,{\sc ii}}}
\newcommand{\siii}{\mbox{Si\,{\sc ii}}}
\newcommand{\mgi}{\mbox{Mg\,{\sc i}}}
\newcommand{\hi}{\mbox{H\,{\sc i}}}
\newcommand{\hii}{\mbox{H\,{\sc ii}}}
\newcommand{\hh}{\mbox{H}$_2$}
\newcommand{\niii}{\mbox{Ni\,{\sc ii}}}
\newcommand{\alii}{\mbox{Al\,{\sc ii}}}
\newcommand{\aliii}{\mbox{Al\,{\sc iii}}}
\newcommand{\kri}{\mbox{Kr\,{\sc i}}}
\newcommand{\nii}{\mbox{N\,{\sc ii}}}
\newcommand{\cii}{\mbox{C\,{\sc ii}}}
\newcommand{\civ}{\mbox{C\,{\sc iv}}}
\newcommand{\ci}{\mbox{C\,{\sc i}}}
\newcommand{\oi}{\mbox{O\,{\sc i}}}
\newcommand{\coii}{\mbox{Co\,{\sc ii}}}
\newcommand{\caii}{\mbox{Ca\,{\sc ii}}}

\title{Large Metallicity Variations\\in the Galactic Interstellar Medium}

\author{Annalisa De Cia$^{1*}$, Edward B. Jenkins$^2$, Andrew J. Fox$^3$, C\'edric Ledoux$^4$, \and Tanita Ramburuth-Hurt$^1$, Christina Konstantopoulou$^1$, \and Patrick Petitjean$^5$, Jens-Kristian Krogager$^1$}

\begin{document}

\date{} 

\maketitle

\begin{enumerate}

 \item Department of Astronomy, University of Geneva, Chemin Pegasi 51, 1290 Versoix, Switzerland
 \item Princeton University Observatory, Princeton, NJ 08544-1001, USA
 \item AURA for ESA, 3700 San Martin Drive Space Telescope Science Institute, Baltimore, MD 21218, USA
 \item European Southern Observatory, Alonso de C\'ordova 3107, Casilla 19001, Vitacura, Santiago, Chile
 \item Institut d'Astrophysique de Paris, Sorbonne Universit\'es \& CNRS, 98bis Boulevard Arago, 75014 Paris, France
 \end{enumerate}

\onehalfspacing
\textbf{The Interstellar Medium (ISM) comprises gases at different temperatures and densities, including ionized, atomic, molecular species, and dust particles \cite{Draine11}. The neutral ISM is dominated by neutral hydrogen \cite{Viegas95} and has ionization fractions up to 8\% \cite{Jenkins13}. The concentration of chemical elements heavier than helium (metallicity) spans orders of magnitudes in Galactic stars \cite{McWilliam97}, because they formed at different times. Instead, the gas in the Solar vicinity is assumed to be well mixed and have Solar metallicity in traditional chemical evolution models \cite{Matteucci12}. The ISM chemical abundances can be accurately measured with UV absorption-line spectroscopy. However, the effects of dust depletion \cite{Field74,Savage96,Jenkins09,DeCia16}, which removes part of the metals from the observable gaseous phase and incorporates it into solid grains, have prevented, until recently, a deeper investigation of the ISM metallicity. Here we report the dust-corrected metallicity of the neutral ISM measured towards 25 stars in our Galaxy. We find large variations in metallicity over a factor of 10 (with an average $55\pm7$\% Solar and standard deviation 0.28~dex) and including many regions of low metallicity, down to $\mathbf{\sim17}$\% Solar and possibly below. Pristine gas falling onto the disk in the form of high-velocity clouds can cause the observed chemical inhomogeneities on scales of tens of pc. Our results suggest that this low-metallicity accreting gas does not efficiently mix into the ISM, which may help us understand metallicity deviations in nearby coeval stars.\\}
\vspace{1cm}

\begin{flushleft}
Note: This version of the article has been accepted for publication on Nature, after peer review, but is not the Version of Record and does not reflect post-acceptance improvements, or any corrections.  The Version of Record is available at \href{http://dx.doi.org/10.1038/s41586-021-03780-0}{http://dx.doi.org/10.1038/s41586-021-03780-0}. Use of this version is subject to the following policies:  \href{https://www.springernature.com/gp/open-research/policies/accepted-manuscript-terms}{https://www.springernature.com/gp/open-research/policies/accepted-manuscript-terms}.\\
Note: An Addendum to this article is appended at the end of this document and is published here: \\ \href{https://www.nature.com/articles/s41586-022-04811-0}{https://www.nature.com/articles/s41586-022-04811-0}.
\end{flushleft}

\begin{figure}
\centering
\includegraphics[width=183mm]{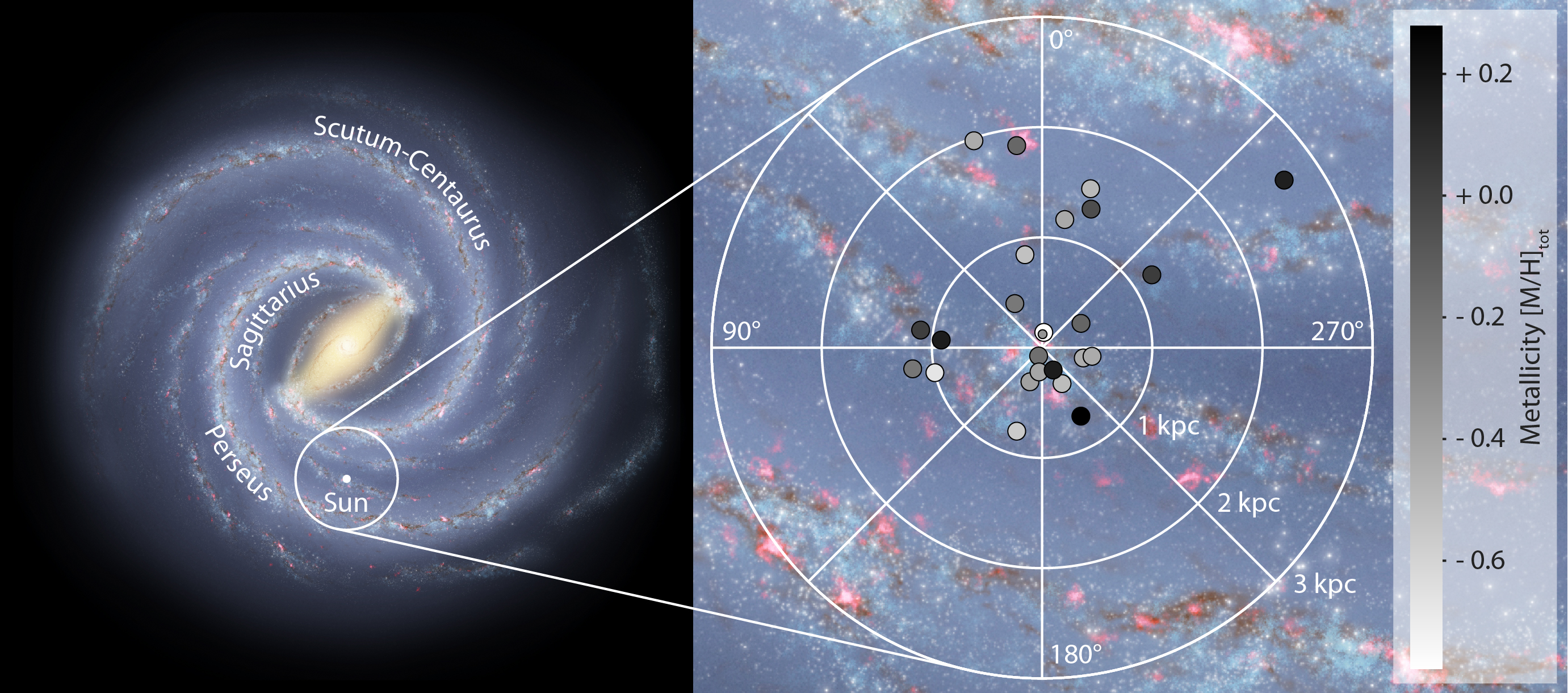}
\caption{\textbf{Location of our targets on the Galactic plane.} Left panel: artistic impression of the Galaxy, face-on, courtesy NASA/JPL-Caltech/R. Hurt (SSC/Caltech). Right panel: the location of our targets is marked on the same illustration of the Galaxy, but zoomed-in on the star-forming spiral arms in the Solar neighborhood. The metallicity of the neutral gas along these lines of sight (Extended Data Table \ref{table met results}) is highlighted with the gray scale. 1~kpc $\sim3.09 \times 10^{19}$~ m.}
\label{fig map}
\end{figure}

We analyze Hubble Space Telescope (HST) Space Telescope Imaging Spectrograph (STIS) near-UV spectra of 25 bright Type O and B stars in the Galaxy. When available, we include archival Very Large Telescope (VLT) Ultraviolet and Visual Echelle Spectrograph (UVES) high-resolution optical spectra of these targets. The selection of the 25 stars in our sample, as well as the data collection and handling, is described in the Methods and the sample properties are listed in Extended Data Table \ref{table sample}. The locations of our targets on the Galactic plane are visualized in Fig. \ref{fig map}. 

We identify and analyze the absorption lines from \mgi, \alii, \siii, \crii, \feii, \coii, \niii, \znii, and \tiii{}. A selection of absorption lines in the spectra of our targets are shown in Extended Data Figure \ref{fig lines}. The column density measurements are explained in the Methods and reported in Extended Data Table \ref{table cold}. We report $3\,\sigma$ significance levels for the quoted limits, and $1\,\sigma$ for the quoted errors, unless otherwise stated.

We determine the neutral gas metallicity using two independent approaches and with different assumptions, which we call the ``relative'' method and the ``$F*$'' method. Both methods aim at characterizing the strength of dust depletion along Galactic lines of sight, and are described in the Methods (Equations \ref{eq [M/H]} to \ref{eq F*}). We use both methods to cross-check our results. In brief, the relative and $F*$ methods use either the gas-phase relative abundances or absolute abundances to estimate by how much the observations are affected by dust depletion based on the empirical relations of \cite{DeCia16} or \cite{Jenkins09}, respectively. Each method uses a specific parameter to estimate the overall amount of dust depletion in a system, either [Zn/Fe]$_{\rm fit}$ or $F*$, as explained in the Methods.

\begin{figure}
\centering
\includegraphics[width=12cm]{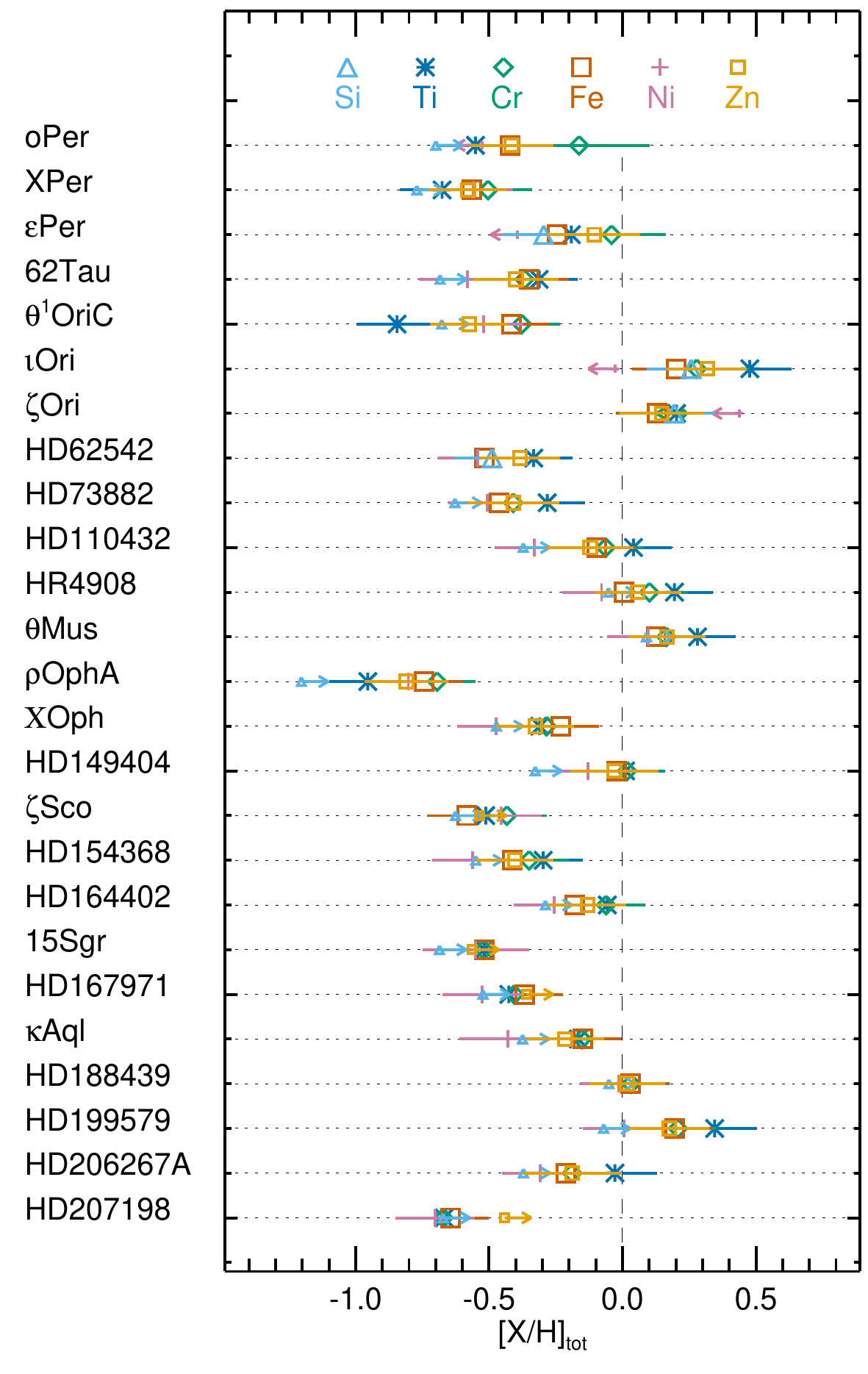}
\caption{\textbf{Dust-corrected abundances in the neutral ISM.} Dust-corrected abundances of Si (light-blue triangles), Ti (blue stars), Cr (green diamonds), Fe (vermilion squares), Ni (pink crosses), and Zn (yellow squares) using the relative method. Stars are labeled in order of HD number (Table \ref{table sample}). The error bars show the $1\,\sigma$ uncertainties.}
\label{fig total xh}
\end{figure}

\begin{figure}
\centering
\includegraphics[width=\textwidth]{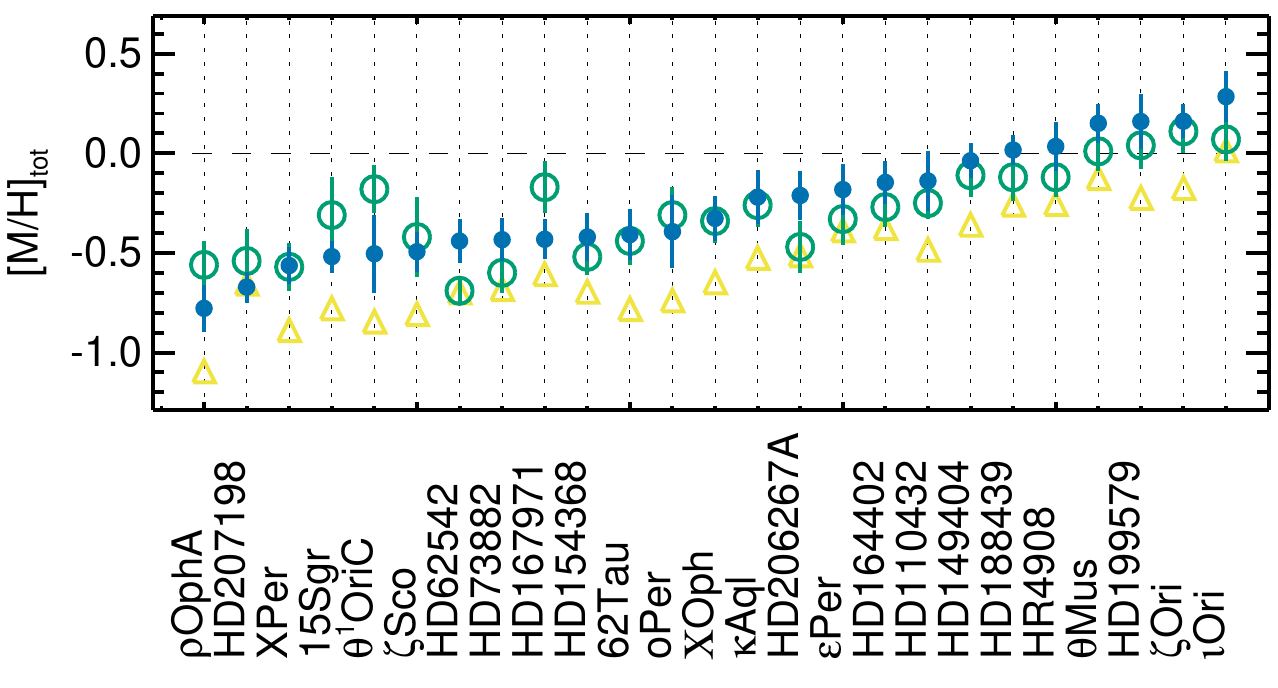}
\caption{\textbf{Metallicities in the neutral ISM.} Dust-corrected metallicities are marked with circles: the filled blue circles are derived with the relative method and the open green circles with the $F*$ method. The open yellow triangles show the observed [Zn/H], and thus represent a lower limit to the true metallicity. Stars are ordered with increasing metallicity. In our sample, 15 out of 25 lines of sight are more than 3 sigma away from Solar metallicity (horizontal dashed line). The error bars show the $1\,\sigma$ uncertainties.}
\label{fig met tot}
\end{figure}

 The individual dust-corrected abundances derived with the relative method are shown in Fig. \ref{fig total xh}, for individual metals. Figure \ref{fig met tot} shows the total metallicities towards the 25 lines of sight in our sample, which are listed in Extended Data Table \ref{table met results} and derived with the relative and $F*$ methods from the fit to the data in Extended Data Figures \ref{fig ZnFe fit} and \ref{fig F* fits}, respectively. We measure total metallicities [M/H]$_{\rm tot}$ ranging between $-0.76$~dex and $+0.26$~dex, where about two thirds of our sample show sub-Solar metallicities. The average metallicity in our sample is $-0.26\pm0.06$~dex (i.e. $55\pm7$\% Solar) with a standard deviation of 0.28~dex. The most striking result is that the maximum variations between lines of sight are more than an order of magnitude, mostly sub-Solar. In the Methods we test this result for different assumptions. The result holds regardless of the assumptions. It is possible that the low metallicities that we measure represent a mix of two different gases, namely a nearly-Solar ISM component with high depletion levels with the injection of significant amounts of pristine gas with zero depletion and metallicities even lower than what we measure (see Methods).

In addition, for some cases where we measure a low metallicity, the most volatile elements (e.g. O, C, Kr) show some disagreement with the mildly-depleted elements (e.g. Zn) and the more refractory elements. This is well explained by the presence of inhomogeneities in the ISM, where nearly-Solar metallicity gas is mixed with large amounts of pristine gas with zero depletion and low metallicity (e.g. 10\% Solar or lower), as we discuss in the Methods. 


While the metallicity of the Sun is $\sim0.2$~dex higher than the expected metallicity of \hii{} regions at a Galactic radius of 8.2~kpc \cite{Arellano20}, this is not enough to explain our results. Local chemical dilution or enrichment in the ISM could be caused by infalling gas or star formation, respectively, but the survival of such chemical inhomogeneities is not straightforward. \cite{Edmunds75} found that chemical enrichment due to Supernovae can induce only small metallicity deviations of $Z\lesssim10^{-4}$ between waves of star formation, after which Galactic rotation can mix the gas in a volume of $\sim0.1$~kpc${^3}$ within a timescale of $2.8\times 10^8$~yr. On the other hand, dilution of disk metallicities to lower values could be caused by accretion of pristine gas, and these inhomogeneities could survive. Chemical evolution models indicate that low-metallicity gas infalling from the halo is necessary to fuel star formation and reproduce stellar abundance patterns \cite{Tosi88,Chiappini97}. Moreover, the large dispersion of the stellar age-metallicity relation \cite{Edvardsson93} suggests a chemically inhomogeneous disk \cite{McWilliam97}. Episodic gas infall could produce pockets of low-metallicity gas and be responsible for the stellar age-metallicity scatter \cite{Pilyugin96}. Analytical inhomogeneous chemical evolution models, which include non-instantaneous mixing of both infalling low-metallicity gas and enriched gas from star formation, showed that significant chemical inhomogeneities due to gas infall (or star formation) will arise if the mass of the new gas cloud is at least $\sim1/20$ times of the overall mass of the gas it is mixing into \cite{White83}. Simulations of SN-driven metal enrichment and ISM mixing suggest that sub-kpc scale inhomogeneities in the ISM should survive at least on timescales of the order of 350 Myr \cite{deAvillez02}. Small sub-parsec-scale variations have been observed in the ISM \cite{Andrews01,Nasoudi-Shoar10}.

Both observations and simulations indicate that gas infall onto the Galaxy halo is metal-poor, with high-velocity cloud (HVC) metallicities observed in the range ~0.1-1 solar \cite{Fox17} and cosmological simulations predicting infalling gas to have metallicities as low as 0.01 solar \cite{Wright21}. More detailed zoom-in simulations show that infalling HVCs can fragment to scales below ~30 pc and disperse on timescales of 10s to 100s of Myr as they mix into the surrounding medium \cite{Gritton14,Heitsch09}. Metallicity inhomogeneities can therefore be smoothed out on these timescales. HVC clouds tend to have individual \hi{} masses of a few $10^5M_\odot$ to $10^6M_\odot$, for a total $M(\mbox{\hi{}})_{\rm tot}\sim10^7M_\odot$ for our Galaxy \cite{Fox17}. Intermediate velocity clouds (IVCs), which tend to have higher metallicities (~Solar) and smaller distances ($<1$--2~kpc) than HVCs \cite{Putman12,Richter17}, may also contribute to the dilution or enrichment of the ISM.

The rate of gas accretion on the Galaxy disk currently measured (0.1--1.4 $M_\odot$/yr, \cite{Lehner11,Fox19}) is a factor of $\sim100$ larger than the minimum required to sustain the formation and survival of chemical inhomogeneities, as we estimate in the Methods. Therefore not only could pockets of low-metallicity/pristine gas exist, but they could in fact be very common. Here we found that two thirds of our sample showed sub-Solar metallicity, but these measurements are integrated along the lines of sight, so that more pockets of lower metallicity gas could exist. The minimum physical scale of the metallicity variations that we observe is of the order of tens of pc, and possibly down to a few pc, see Methods. Our methodology opens up the possibility of comparing for the first time the metallicity of the neutral gas with the metallicity of stars and their \hii{} regions, although this is not a straightforward comparison, as we discuss in the Methods.

Finally, we find no positive correlations between the total metallicities and the $E(B-V)$, meaning that we do not observe higher reddening towards regions of higher metallicity. We do not find a significant correlation between $R_V$ and metallicity, but we note that $\theta^1$~Ori~C and $\rho$~Oph~A have high values of $R_V$ (Table \ref{table sample}). The dilution of the ISM with lower-metallicity gas may play a role in explaining these effects. We do not observe any significant signs of a radial metallicity gradient in the gaseous disk, as shown in Extended Data Figure \ref{fig met radial}. The observed metallicity gradients in the Galaxy have slopes of up to $\sim-0.07$~dex/kpc, from mesurements of stars \cite{Cheng12} and \hii{} regions \cite{Arellano20}, but they are probed over several kpc. Our measurements cover a much narrower range (3 kpc) and with smaller statistics. A larger sample is needed for further conclusions on a potential gradient. We also do not observe any trend of metallicity with galactic height, as shown in Extended Data Figure \ref{fig met radial}. Thus, it is unlikely that the accreting gas is evenly distributed above the disk, but it is consistent with being clumpy, as in HVCs or tidal streams. On the other hand, our targets mostly lie within the young thin disk, probing small height scales. The global effects in chemical enrichment of inflowing and outflowing gas above the galaxy disks can be more evident at circumgalactic scales \cite{Wendt20}.

We conclude that we have measured large local variations of metallicity in the neutral ISM in our Galaxy likely due to accretion of low-metallicity gas. Thus, we recommend changing the common assumptions that the gas in galaxies is well mixed and the gas in the Galaxy has Solar metallicity in the Solar vicinity, which are widely used both in observational and theoretical works, and in particular for the study of the chemical evolution of the Galaxy. Our findings that the gas is inhomogeneously distributed (not only chemically) indicate that the gas mixing is more inefficient than previously thought. One of the potential causes could be the fact that the different phases involved in the mixing have widely different kinematics and different physical conditions. In addition, substantial sustained gas inflow may contribute to the clumpiness and turbulent nature of the disk in addition to gravitational instabilities.

\clearpage

\newpage

\newpage

\clearpage

\section*{Methods}

\vspace{0.5cm}

\begin{flushleft}
\textbf{Sample selection and data collection}\\
\end{flushleft}
We select 25 stars and observe them as part of the cycle 25 HST program ID 15335 (PI De Cia). The sample was selected to include hot stars (spectral Type O or B) which have line-of-sight measurements of \hi{}, \hh{}, and \tiii{} in the literature \cite{Savage77,Diplas94,Welty10}, high-enough rotational velocities ($\gtrsim 50$ km~s$^{-1}$) to allow us to disentangle between stellar and ISM feature in the spectra, and fairly diverse values of dust reddening and Galactic longitude. The distances $r$ are taken from the Gaia DR2 archive \cite{Gaia16, Gaia18}. The values of \hi{} and \hh{} are taken from \cite{Diplas94} and \cite{Savage77}, respectively, as also collected by \cite{Welty10} for most of our targets, and assuming their average uncertainties of 0.1~dex. For HD~62542 we adopt the column densities of atomic and molecular hydrogen from \cite{Welty20}. The main characteristics of the 25 selected stars are reported in Extended Data Table \ref{table sample}. 

The reddening $E(B-V)$ of the stars in our sample is taken from \cite{Diplas94} and spreads between 0.07 and 1.08 mag. The values of the optical total-to-selective extinction ratio, $R_V = A_V / E(B-V)$, from the maps of \cite{Valencic04} show variations between 2.6 and 5.7. The positions of the stars are distributed in Galactic longitude, their distances are within 2.7~kpc from the Sun, and they lie within 0.3~kpc from the disk plane. All lines of sight have total hydrogen column densities ($N({\rm H}) = N(\mbox{\hi{}}) + 2 \times N(\mbox{\hh{}})$) that are high enough to shield the gas against photoionization \cite{Viegas95}. We further discuss potential ionization effects below.

All stars are observed with the same HST/STIS NUV echelle setting E230M (at a central wavelength setting of 1978 \AA{}), which covers absorption lines between $\sim1605$ and $\sim2832$ \AA{} with a resolving power  $R = \Delta \lambda / \lambda \sim 30,000$. 

We further collected all the available reduced VLT/UVES spectra in the ESO Science Archive covering \tiii{} $\lambda$ 3230, 3242, 3384 lines for our sample, which are available for 16 of our targets and have $R \sim 71,050$. We stack the spectra for each star to maximize the S/N, and measure the \tiii{} column densities in the same way as for the other metals, which are reported in Extended Data Table \ref{table cold}. In the event that no data that cover \tiii{} are available, we adopt the \tiii{} column densities of \cite{Welty10}.   

For the star HD~62542 we include the column densities that were recently measured by \cite{Welty20} from HST data at a higher spectral resolution. Given the better quality of these data, we adopt their column densities for both of their velocity components summed together for our analysis.  

\vspace{0.5cm}

\begin{flushleft}
\textbf{Analysis of the absorption features and column density determination.}\\
\end{flushleft}

We identify and analyze the following absorption lines: \mgi{} $\lambda$ 1827 and 2026; \alii{} $\lambda$ 1670; \siii{} $\lambda$ 1808; \tiii{} $\lambda$ 3230, 3242, and 3384; \crii{} $\lambda$ 2056, 2062, and 2066; \feii{} $\lambda$ 2344, 2260, and 2249; \coii{} $\lambda$ 2012 and 1941 (although we did not use the column densities of \coii{} in our analysis); \niii{} $\lambda$ 1741, 1709, and 1751; \znii{} $\lambda$ 2026 and 2062. The line profiles of the \znii{} $\lambda$ 2026, \crii $\lambda$ 2056, and \feii{} $\lambda$ 2260 transitions are shown in Extended Data Figures, \ref{fig lines}. The velocity extremes over which we performed absorption line measurements appear to be well defined by the maximum extent of the very strong \feii{} $\lambda$ 2344 transition. We used the most recent oscillator strengths ($f$-values), listed in Extended Data Table \ref{table lines}. We determined the column densities by integrating the apparent optical depths (AOD) \cite{Savage91} over all velocities between these extremes, regardless of the appearance of the line in question. This procedure guaranteed that we did not overlook weak parts of an absorption buried in the noise, and also prevented us from defining a continuum level partly within the extent of such absorption. We defined continuum levels from best-fitting Legendre polynomials to fluxes on either side of the absorption profiles. Some lines appeared to be saturated or nearly so; in such cases the AOD method can underestimate the true column density. When the lowest part of a feature seemed close to the zero intensity level but still had a pointed appearance, we considered saturation to be taking place and declared the measurement as a lower limit. The \feii{} $\lambda$ 2344 line was often heavily saturated, but weaker \feii{} lines ($\lambda$ 2260, 2249) were well constrained. In most circumstances, the two lines of Zn II did not appear to be appreciably saturated, but the fact that the stronger $\lambda$ 2026 line consistently gave a column density that was lower than the weaker $\lambda$ 2062 line indicated that some unresolved saturation was still creating column density outcomes that were below the true ones. To overcome this problem, we invoked the scheme proposed by \cite{Jenkins96} that applies a correction to arrive at a more accurate column density. The contribution of nearby \mgi{} $\lambda$ 2026 and \crii{} $\lambda$ 2062 features were taken into account for the calculation of the \znii{} column density. We estimated errors for the column densities from the effects of three different sources: (1) noise in the absorption profile, (2) errors in defining the continuum level, and (3) uncertainties in the transition f-values, all of which were combined in quadrature. Continuum placement errors can have a large influence in the uncertainties of weak lines. We evaluated the expected deviations produced by such errors by remeasuring the AODs at the lower and upper bounds for the continua, which were derived from the expected formal uncertainties in the polynomial coefficients of the fits as described by \cite{Sembach92}. We multiplied these coefficient uncertainties by 2 in order to make approximate allowances for additional deviations that might arise from some freedom in assigning the most appropriate order for the polynomial. We considered a measurement to be marginal if the equivalent width outcome was less than the 2$\sigma$ level of uncertainty from noise and continuum placement. For weak lines below this uncertainty threshold, we specified an upper limit for the column density. Details of how we calculated these 1$\sigma$ upper limits are given in Appendix D of \cite{Bowen08}. If the strongest transition yielded an upper limit or a very marginal detection, no attempt was made to measure considerably weaker ones except when a weaker line was in a much better part of the spectrum (higher signal-to-noise ratio or a more easily defined continuum).

We use a linear unit for the column densities $N$ in terms of ions cm$^{-2}$. We refer to relative abundances of elements $X$ and $Y$ as $\left[X/Y\right] \equiv \log{\frac{N(X)}{N(Y)}} - \log{\frac{N(X)_\odot}{N(Y)_\odot}}$, where reference Solar abundances are listed in \cite{DeCia16}.

\vspace{0.5cm}

\begin{flushleft}
\textbf{Potential saturation effects and comparison with the Voigt-profile fit method}\\
\end{flushleft}
Saturation effects do not play an important role for the results presented in this paper. If present, any potential saturation effect must be small, and in particular much smaller compared to the strong effects of dust depletion. This can be immediately appreciated by the small deviations from the linear fits in Extended Data Figures \ref{fig ZnFe fit} and \ref{fig F* fits}. In addition, if saturation of \znii{} would be an issue, we should see more deviations in the dust-corrected abundances of Zn at Solar metallicity (Figure \ref{fig total xh}), which are not observed.

Nevertheless, we tested the robustness of the column density determinations with the AOD method with an independent method, the Voigt-profile fit, which decomposes and models the line profiles in their individual velocity components and fits all transitions simultaneously. The Voigt-profile fit can use narrow $b$-values to account for saturation. These tests are aimed at further assessing potential saturation of \znii{} absorption lines. Saturation for other ions considered here is less probable because of the availability of weaker absorption lines to measure. Using the VoigtFit software \cite{Krogager18}, we model lines of \znii{} and \crii{} towards the eight targets that have the strongest \znii{} absorption and show potential for saturation (Extended Data Figure \ref{fig lines}), and with a column density of \znii{} constrained with the AOD method ( Extended Data Table \ref{table cold}), namely: $\theta^1$~Ori~C, HD~73882, HR~4908, $\theta$~Mus, HD~149404, HD~154368, HD~199579, and HD~206267. This includes the most troublesome target, $\theta^1$~Ori~C, which we discuss below.

For the seven out of the eight lines of sight tested, with the exception of $\theta^1$~Ori~C, the column densities of \crii{} measured with the AOD and Voigt-profile fit are in excellent agreement (mostly within 0.03~dex and consistent within the errors) and the column densities of \znii{} measured with the Voigt-profile fit differ by $+0.03$, $+0.22$, $-0.1$, $-0.1$, $-0.02$, $-0.08$, and $+0.06$~dex with respect to the AOD results, corresponding to $0.2$, $1.8$, $2.6$, $1.0$, $0.3$, $1.0$, and $0.5$ $\sigma$ (Z-test) for HD~73882, HR~4908, $\theta$~Mus,  HD~149404, HD~154368, HD~199579, and HD~206267, respectively. These are mostly typical values for the comparison between AOD and the Voigt-profile fit. The larger discrepancies are measured for HR~4908 and $\theta$~Mus, towards which we estimated near-Solar metallicity (Extended Data Table \ref{table met results}).

A different picture arose for $\theta^1$~Ori~C, for which we measured significant differences of $+0.33$ and $+0.45$~dex with respect to the AOD results, corresponding to 7.4 and 3.6 $\sigma$ (Z-test), for \znii{}, and \crii{}, respectively. However, the Voigt-profile fit of the HST/STIS data was not well constrained for this system. This is likely caused by the extreme complexity of this line of sight \cite{Price01}, which cannot adequately be decomposed with the sampling of 5~km/s per pixel of the HST/STIS data. On the other hand, with a Voigt-profile fit to the VLT/UVES data we obtained consistent measurements (within 1~$\sigma$) with the AOD \tiii{} results and the results of \cite{Price01} for \caii{}. The complexity of this line of sight may also explain the discrepancies among the different metals that we observe in Extended Data Figures \ref{fig ZnFe fit} and \ref{fig F* fits}. In the latter, the \znii{} column is significantly below the linear fit, so that an increase in \znii{} column would not affect heavily the result. These discrepancies are the likely cause of the differences in metallicities that we measured towards this line of sight with the relative and $F*$ method (Extended Data Table \ref{table met results}).

Overall, our results hold regardless of whether the Voigt-profile fit or AOD method is used for estimating the column densities.

\vspace{0.5cm}

\begin{flushleft}
\textbf{Ionization effects on column density measurements.}\\
\end{flushleft}
Due to the background radiation field in the ISM, the atoms in the gas phase are found in various ionization states (e.g., \cii{} / \civ{}, \alii{} / \aliii{}). In the neutral medium, the singly ionized metals are the dominant species due to their ionization potentials relative to that of \hi{}. In order to obtain metallicities, we assume that other ionization states are negligible such that: $N(X) / N(\mbox{H}) = \sum_i N(X_i) / ( N(\mbox{\hi{}}) + N(\mbox{\hii{}}) )$, where the summation is over all ionization states of a given metal, $X$. However, if the total integrated absorption line arises from many unresolved components with varying physical properties, it is possible that this assumption does not hold for individual components. Such ionization corrections could therefore affect our conclusions about a difference in metal enrichment being the driver of the variations in the depletion sequences.

In all but three lines of sight, we detect absorption from neutral carbon (with an ionization potential lower than that of \hi{}), which indicates that the gas phase must be highly shielded by \hi{}. For the three cases where no clear \ci{} absorption is seen ($\epsilon$~Per, $\iota$~Ori, $\zeta$~Ori~A), the non-detections are consistent with the overall low optical depth of other lines. The \ci{} line profiles resemble those of the singly ionized metals indicating that they arise from the same gas phase. These facts provide strong evidence that the gas is effectively shielded; Hence, ionization corrections can be neglected.

If ionization effects were the culprit of the observed deviations in the relative abundances, and not differential dust depletion, then the oxygen abundance should not follow those of the other volatile elements given the tight relationship between \hi{} and \oi{} (due to charge exchange reactions). However, in the cases where we can constrain the oxygen abundance, we see that it does indeed show the same behaviour as the heavier volatile elements. This further bolsters our conclusion that ionization effects are negligible.

\vspace{0.5cm}
 
\begin{flushleft}
\textbf{The relative and $F*$ methods to measure the ISM metallicity}\\
\end{flushleft}
To measure the ISM metallicity it is essential to quantify the amount of metals that are missing from the observable gas-phase but instead are incorporated into dust grains, which is the phenomenon of dust depletion \cite{Field74,Phillips82,Jenkins86,Savage96,Jenkins09,DeCia16,RomanDuval21}. The dust-corrected (total of gas and dust) abundances can defined as 
\begin{equation}
[X/{\rm H}]_{\rm tot} = [X/{\rm H}]- \delta_X \mbox{,}
\label{eq [M/H]}
\end{equation}
where $[X/{\rm H}]$ is the observed abundance of metal $X$ and $\delta_X$ is its depletion in dust (which is a negative term in this classical notation). Here no other effects such as nucleosynthesis or ionization are taken into account. The depletion $\delta_X$ is linearly proportional to the overall strength of depletion \cite{Jenkins09,DeCia16,RomanDuval21}. The overall strength of depletion can be represented in different ways, for example from the observed relative abundances, like in the ``relative method'', or with a specific parameter $F*$, like in the ``$F*$ method''.
 
The relative method was first introduced by \cite{DeCia16} who compared galactic and extragalactic observations. To estimate the overall strength of depletion, this method uses any relative abundance [$X/Y$] where $X$ and $Y$ follow each other in nucleosynthesis, but have very different refractory properties. A dust tracer can be [Zn/Fe], or other relative abundances such as [Si/Ti] or [O/Si]. \cite{DeCia18} discusses in more detail why [Zn/Fe] is a reliable dust tracer in the metallicities ranges considered here. The depletion of element $X$, $\delta_X$, is obtained from the observed correlation between [$X/Y$] (with $Y$ non refractory such as S or P - here we use Zn) and the dust tracer [Zn/Fe] (or any other dust tracer). The dependency of [$X$/Zn] on the depletion of Zn can be removed by assuming a certain slope for the expected depletion of Zn with the dust tracer, $B_{\delta_{\rm Zn}}$. Then $\delta_X$ can be derived as follows:
\begin{equation}
\delta_X = A2_X + B2_X \times [{\rm Zn/Fe}]\mbox{,}
\label{eq delta}
\end{equation}
where $A2_X$ and $B2_X$ are reported in Extended Data Table \ref{table AB}. The simplest version of the relative method uses directly the observed [Zn/Fe] to estimate the dust depletion of the different metals. However, the information on all metals can be used simultaneously to determine the overall strength of the dust depletion with the parameter [Zn/Fe]$_{\rm fit}$, which is equivalent to [Zn/Fe] but is derived from the information of all metals. Merging Eqs. \ref{eq [M/H]} and \ref{eq delta}, and using the basic definition of [$X$/H], it is possible to find the dust-corrected metallicity and overall strength of depletion from the observed metal column densities, through a fit of the linear relation
\begin{equation}
y = a + bx \mbox{,}
\label{eq xy}
\end{equation}
where
\begin{equation}
a = [{\rm M/H}]_{\rm tot}\mbox{,}
\label{eq a}
\end{equation}
\begin{equation}
b = [{\rm Zn/Fe}]_{\rm fit}\mbox{,}
\label{eq b}
\end{equation}
\begin{equation}
x = B2_X\mbox{,}
\label{eq x}
\end{equation}
\begin{equation}
y = \log N(X) - X_\odot + 12. - A2_X - \log N({\rm H})
\label{eq y}
\end{equation}
and considering the uncertainties on both $x$ and $y$. $x$ and $y$ represent the depletion data for different metals, and we fit a linear relation to this data to find the parameters $a$ and $b$, which are unique to each system. In this way, the $y$-intercept (at $x = 0$) of the fitted relation gives the total metallicity, [M/H]$_{\rm tot}$, and its slope gives the overall strength of depletion, [Zn/Fe]$_{\rm fit}$. Extended Data Figure \ref{fig ZnFe fit} illustrates this procedure for our sample. The fitted and observed [Zn/Fe] often agree well, see Table \ref{table met results}.

Changes in relative abundances could in principle be caused by nucleosynthesis processes or other reasons. In particular, deviations from the curves of Extended Data Figures \ref{fig ZnFe fit} and \ref{fig F* fits} could in principle be caused by nucleosynthesis. However, the relative abundance patterns that we observe here mostly follow the refractory properties of the metals, rather than their nucleosynthesis origin, and are therefore caused purely by dust depletion. We do not expect the infalling gas to have been enriched in specific metals or show or have any peculiar abundances due to different nucleosynthetic history. Alternatively, one could have the opportunity to investigate such peculiar abundances by studying eventual deviations from the depletion patterns.

The choice of $B1_{\delta_{\rm Zn}}$ is the main assumption of the relative method used here to find dust-currected metallicities. First, it is reasonable to assume that $\delta_{\rm Zn}$ correlates linearly with [Zn/Fe] or other dust tracers, because this is observed for all other metals \cite{DeCia16, Jenkins09}. Moreover, we know that at [Zn/Fe] $=0$ there is no dust depletion of non-carbonaeous species and we can safely assume no depletion of Zn, i.e. $\delta_{\rm Zn}=0$ at [Zn/Fe] $=0$. Thus, the main free parameter in the assumption of the Zn depletion is the slope of the relation of $\delta_{\rm Zn}$ with [Zn/Fe], $B1_{\delta_{\rm Zn}}$. In this paper we assume $B1_{\delta_{\rm Zn}} = - 0.27\pm 0.03$ derived by \cite{DeCia16}. We conservatively test our results for different slope assumptions, as described below, to ensure that our ultimate results are not affected. Notably, the  $A2_X$ and $B2_X$ slopes for Ti are not well constrained to date, and this is a weakness of the relative method. However, this does not affect the overall results of this paper, which can mostly be derived already from the basic version of the relative method, using only the observed [Zn/Fe] and regardless of Ti, with the exception of those systems where $N({\rm Zn})$ is not constrained.

The $F*$ method, first developed by \cite{Jenkins09}, characterizes the dust depletion in Galactic clouds, by correlating all the observed abundances and minimizing the residuals with respect to a common factor, $F*$. This factor represents the overall strength of dust depletion in individual lines of sight. This method assumes that the underlying metallicity is Solar. In an analogous way as described above, the metallicities are found by fitting the linear relation
\begin{equation}
y = \log N(X) - \log N({\rm H}) - X_\odot - B_X + A_X \times z_X  \mbox{,}
\label{eq F*}
\end{equation}
where $A_X$, $B_X$, and $z_X$ coefficients are determined and listed in \cite{Jenkins09}. The definition of $y$ differs by $- \log N({\rm H})$ with respect to \cite{Jenkins09}. The fit of this $y = a + bx$ relation, where $x = A_X$ yields a $y$ intercept at $x = 0$ equal to [M/H]$_{\rm tot}$, and $b = F*$.

Extended Data Figures \ref{fig ZnFe fit} and \ref{fig F* fits} show how the relative and $F*$ methods fit the overall strength of dust depletion ([Zn/Fe]$_{\rm fit}$ or $F*$) and total (dust-corrected) metallicity [M/H]$_{\rm tot}$ to the the observed column densities. There is an overall good agreement between the total metallicity resulting from the relative method and the $F*$ method (open green circles), with a few notable exceptions, i.e. for $\theta^1$~Ori~C and HD~62542. These are cases of peculiar abundances in particular of Ti, and possibly Zn, that make the depletion patterns vary considerably from the norm, as visible in Extended Data Figures \ref{fig ZnFe fit} and \ref{fig F* fits}. HD~62542 is a case with strong ISM inhomogeneities along the line of sight, with one cloud having a much stronger depletion than the others \cite{Welty20}, and maybe also different metallicities.

The main assumption in the original establishment of the $F*$ method is that the metallicity of the gas is Solar, and this assumption was used to compute the $A_X$, $B_X$, and $z_X$ coefficients \cite{Jenkins09}. That is, the individual observed [$X$/H] used to determine $A_X$ and $B_X$ are treated as pure signs of dust depletion, not including any potential variation in metallicity. However, the $A_X$ and $B_X$ coefficient determination could be altered, in case of significant deviations from the Solar metallicity, such as potential several low-metallicity clouds. 

Both the relative and $F*$ methods are sensitive to the effect of potential ISM dishomogeneities along the line of sight, which we discuss below.

\vspace{0.5cm}

\begin{flushleft}
\textbf{Testing the assumptions of the relative method}\\
\end{flushleft}
We test our results using different assumptions on the slope of the Zn depletion sequence $B_{\delta_{\rm Zn}}$, which is the main assumption of the relative method, for the basic version of the method, i.e. relying on the observed [Zn/Fe] only to characterize the $\delta_X$. In addition to the optimal slope for the depletion of Zn ($B_{\delta_{\rm Zn}}=-0.27$ \cite{DeCia16}), we test for slopes that are two times steeper, two times shallower, and from the relations of \cite{Jenkins09}. \textit{i)} Assuming a steeper slope ($B_{\delta_{\rm Zn}}\equiv-0.54$), the resulting dust-corrected metallicities are 0.3--0.4~dex higher than using the optimal slope. However, such steep depletion of Zn is quite extreme and very difficult to reconcile with the independent work on dust depletions of \cite{Jenkins09} (see Fig. 5 of \cite{DeCia16}). We consider this only as an extreme option of our parameter space to explore the potential impact on our results. \textit{ii)} Assuming a shallower slope ($B_{\delta_{\rm Zn}}\equiv-0.135$), the resulting dust-corrected metallicities are 0.15--0.2~dex lower than using the ``reference'' slope $B_{\delta_{\rm Zn}}$. This assumption is a potentially plausible option. \textit{iii)} Finally, we assume that the depletion of Zn has the distribution of \cite{Jenkins09} and use the (assumed) linear correlation between $F*$ and [Zn/Fe] \cite{DeCia16}, and thus assuming $A_{\delta_{\rm Zn}}=0.785$ and $B_{\delta_{\rm Zn}}=-0.904$. Note that the extrapolation of this correlation to negative $F*$ is not necessarily reliable. Overall, the slopes for the metal depletion of \cite{Jenkins09} are steeper than in \cite{DeCia16}. The metallicities derived with this last assumption are overall similar to those using the optimal slope, with a few local variations towards higher metallicities, in one case up to 0.3~dex. One common feature of our results, regardless of the assumption on the Zn depletion (and even including the less likely assumptions) is a wide spread in metallicities of $\sim1$~dex.

\vspace{0.5cm}

\begin{flushleft}
\textbf{The influence of ISM inhomogeneities along the line of sight}\\
\end{flushleft}
One important concern is the presence of line-of-sight inhomogeneities, and in particular whether our methods of estimating the metallicities could be affected by the ISM being composed of individual clouds with very different depletion strengths and/or metallicities. One example of such inhomogeneities is star HD~62542, for which individual components with very different depletion properties have been observed with higher resolution spectroscopy (see Fig. 6 of \cite{Welty20}). Most \tiii{} line profiles in the stacked UVES spectra in our sample show asymmetry or complexity in the velocity structure, hinting at potentially separate clouds or inhomogeneous ISM along the line of sight.

We test the application of our methods for the case of two individual ISM components with different combinations of amount of gas, depletion, and metallicity. If the two individual components have the same metallicity, the metallicity determination that we would estimate from the combined total column densities (over the whole line profiles) would be accurate, provided we use the $y$ variables as defined in Eqs. \ref{eq y} and \ref{eq F*}. However, if the two ISM components not only have very different depletion strengths, but also very different metallicities, then our estimates of the metallicity using the combined total column densities is in between the metallicities of the two clouds, and mostly closer to the components that carries most gas. For example, if one component has $F*=1$, Solar metallicity, and carries 10\% of the gas, while the second component has $F*=0$, 10\% Solar metallicity, and carries 90\% of the gas, the relative method finds a global metallicity of 18\% Solar.

The injection of pristine gas (metal-poor gas with zero depletion levels) in the ISM can bring strong deviations from a straight-line fit in the $x$ vs $y$ plots, which affects the metallicity determination. If the line-of-sight ISM is composed of two clouds carrying the same amount of gas, the first one with [Zn/Fe] $=1.6$ and Solar metallicity, and the second one with [Zn/Fe] $=0$ and 10\% Solar metallicity, then the relation between the total $x$ and $y$ (from column densities measured over the whole line profiles) in the relative method strongly departs from a straight line, with a curvature where the more volatile elements have higher than expected $y$ values.

In fact, we observe sometimes large discrepancies between the observed abundances of the highly volatile elements (e.g. O, Kr, C, N) taken from the literature (\cite{Jenkins09}, \cite{Jenkins19}, see Extended Data Table \ref{table volatile}) on one hand and on the other hand the observed abundances of the mildly volatile elements (e.g. Zn) and the refractory elements, which is hard to interpret with the classic knowledge of dust depletion. This effect is highlighted in Extended Data Figure \ref{fig F* fits}, and is observed for several lines of sight in our sample, but only among those for which we find low metallicities (namely, $o$~Per, X~Per, 62~Tau, HD~62542, $\rho$~Oph~A, $\chi$~Oph, HD~154368, 15~Sgr, and HD~207198). Classically, this effect was attributed to very high values of $F*$ (1--1.6), but with the difficulty of reconciling the overall depletion patterns. These abundance patterns can also not be attributed to nucleosynthesis effects. In principle, there could be intrinsic differences between O and N because of their different nucleosynthetic origin (primary $\alpha$-capture process or secondary processes). However, only in one case we report measurements of nitrogen only among the volatile elements (HD~110432, Extended Data Figure \ref{fig F* fits}), and otherwise the measurements of N agree with all the other volatile elements. 

Here we suggest that the large differences between the observed abundances of the most volatile elements with respect to Zn and the refractory elements are the effect of a significant injection of pristine gas (with low metallicity, e.g. 10\% Solar, and no dust depletion) to the main ISM component with more typical (Solar) metallicity and depletion levels. In our calculations we assume that the pristine gas with no depletion has 10\% Solar metallicity, but it could have even lower metallicities. Dust depletion is often observed at high redshift in systems with 10\% Solar metallicity \cite{DeCia18b}.

The line of sight towards HD~62542 is a good, and perhaps extreme, example of a case where two (or more) ISM components with very different depletion properties (one with extremely high $F*=1.5$ and the rest with $F*=0.3$, confirmed with higher resolution spectroscopy). But even after taking this into account, the observed abundances do not fit the depletion patterns well \cite{Welty20}. In our analysis we found sub-Solar values of metallicities, although we only considered the total columns measured over the whole line profile. Thus, we speculate that additional ISM mixing effects, and in particular with gas at sub-Solar metallicities, may in fact help reconcile the observations.

Overall, in the case of an inhomogeneous ISM with contributions of gas at different metallicities, our total metallicity estimates likely represent a value in between the lowest and highest metallicities, depending on how much gas the clouds bear. Our results witness significant amounts of low metallicity gas. Some of these could potentially represent a mix between nearly-Solar metallicity and pristine gas, with even lower metallicities than our integrated estimates.

\vspace{0.5cm}

\begin{flushleft}
\textbf{Estimates of the gas accretion rate}\\
\end{flushleft}
We roughly estimate the minimum accretion rate required for clouds of gas not to mix efficiently, which would allow chemical inhomogeneities to survive. We assume that a volume $V_{\rm mix}\sim0.1$~kpc${^3}$ completely mixes within a timescale of $\tau_{\rm mix}=2.8\times 10^8$~yr due to Galactic rotation \cite{Edmunds75}, and calculate the mass of this mixing volume by assuming a number density. For this, we compute the average number densities along the lines of sight in our sample, $n_{\rm{H}} = N({\rm H}) / r$ using only the accurate distance measurements from Gaia DR2, and adopt their mean value $<n_{\rm H}> = 2.3$~cm$^{-3}$. This is a typical value for the warm neutral medium \cite{Draine11}. The mass of the mixing region is then $M_{\rm mix}\sim6.2\times10^6M_\odot$. Given the minimum mass fraction for chemical inhomogeneities to survive (1/20, \cite{White83}), the minimum accreting mass that can lead to a long-lived chemical inhomogeneity is $m_{\rm min, acc}\sim3.1\times 10^5M_\odot$. This is comparable with the mass of a small HVC cloud \cite{Fox17}. Overall, the minimum accretion rate that allows chemical inhomogeneities to survive Galactic rotation is $R_{\rm min, acc}=m_{\rm min, acc} / \tau_{\rm mix} \sim 0.001M_\odot$/yr. This is much smaller than the typical accretion rate in the Galaxy, which is estimated to be around 0.1--1.4 $M_\odot$/yr from UV observations of HVCs \cite{Lehner11,Fox19}, not including the Magellanic Stream. That is, the rate of gas accretion that is currently measured is more than enough (by a factor of $\sim100$) to account for the existence and survival of chemical inhomogeneities. This is true even if we would assume a mean number density of ten times higher, typical of the cold neutral medium, in which case $R_{\rm min, acc}$ would be ten times higher.  

\vspace{0.5cm}

\begin{flushleft}
\textbf{Estimates of the physical scale of the metallicity variations}\\
\end{flushleft}
The physical scale of the metallicity variations cannot be easily estimated, because of the sparsity of our targets, and the fact that we measure the integrated metallicity along their lines of sight. The smallest scales can perhaps be probed in the Orion region, in which we observed metallicity variations towards three targets. The angular separation between $\theta^1$~Ori~C, a likely low-metallicity line of sight, and $\iota$~Ori and $\zeta$~Ori~A is $\sim30.7$\arcminute and $\sim3.7325$\degree, respectively. At the physical distance of $\theta^1$~Ori~C, 373~pc, these corresponds to physical separations of $\sim3.3$ and $\sim24$~pc. Regarding the physical distance to our targets, the closest star  is $\epsilon$~Per at 82~pc and we measure about Solar metallicity for the neutral gas toward this line of sight. Then we measure low metallicity towards the next two closest targets, $\chi$~Oph and $\rho$~Oph~A at 122 and 139~pc, respectively, and with metallicities of 37\% and 13\% the Solar value. While above we conservatively refer to the distances among the stars themselves, the physical sizes of the pockets of lower-metallicity gas and the distances among them could be smaller. All metallicities are listed in Extended Data Table \ref{table met results}. Thus, we roughly estimate the minimum physical scale of the metallicity variations that we observe to be of the order of tens of pc, and possibly down to a few pc. 

\vspace{0.5cm}
     
\begin{flushleft}
 \textbf{Comparing the metallicity of neutral gas, ionized gas, and stars}   
 \end{flushleft} 
An interesting open issue is how the neutral gas metallicity is related to the \hii{} regions and in turn to the stellar metallicities. We are not aware of measurements of stellar metallicities of the stars in our sample. For the case of the Orion OB association, \cite{SimonDiaz10} found that the B-type stars in the Orion OB association, including in the Orion Nebula, have around Solar metallicity. The metallicity of the ionized gas in the Orion nebula \hii{} region is debated, ranging from 1/10 Solar metallicity (for collisionally excited lines, and in general for refractory elements like Fe, Mg, and Si \cite{Rubin93,Garnett95,SimonDiaz11} to slightly supersolar \cite{Esteban04}. \cite{Esteban18} found deviations up to $\sim0.4$~dex (0.2) from the abundances of nitrogen (oxygen) of ionized gas in \hii{} regions in the Galaxy, after accounting for a radial metallicity gradient. \cite{Balser15} and \cite{Arellano20} measured the oxygen metallicity in \hii{} regions in our Galaxy and found metallicity variations over the overall radial gradient of $\sim0.3$~dex over large scales, and smaller along the direction of the long bar. \cite{Wang20} recently found that $\sim0.5$~dex variations of gas-phase metallicity on scales of 100~pc, observed from strong-emission lines diagnostics in Very Large Telescope/Multi Unit Spectroscopic Explorer (MUSE) data of nearby galaxies, positively correlate with the variations in star-formation rates, which they interpret as due to time variations of the star-formation efficiency and including gas-infall in their gas-regulated theoretical models. With similar observational techniques, \cite{Kreckel19} found that the metallicity of \hii{} regions in nearby galaxies show deviations of up to 0.2--0.3~dex from the radial gradients, and these variations decrease at smaller scales \cite{Kreckel20}. 

Comparison with our results is not straightforward. First, \hii{} regions comprise ionized dense gas that is associated with active star formation, while the diffuse neutral gas that we probe in this paper is not necessarily connected to recent star-formation activity. Second, here we probe individual lines of sight, down to very small projected physical scales of a few pc. Such small-scale variations can be observationally smoothed away in integrated measurements of metallicity over larger scales, which is normally the case for nearby galaxies. In addition, \hii{} measurements can also be affected by dust depletion, including for the most commonly used reference element, oxygen.

     
\begin{flushleft}
\onehalfspacing
\textbf{Acknowledgements} \hspace{0.5cm} A.D.C. thanks Cesare Chiosi and the ``Galaxies and the Universe'' group at the University of Geneva for discussions, and Barry Holl for help with navigating the Gaia archive. A.D.C., T.R.H., C.K., and J.K.K. acknowledge support by the Swiss National Science Foundation under grant 185692. Based on observations with the NASA/ESA Hubble Space Telescope obtained at the Space Telescope Science Institute (STScI), which is operated by the Association of Universities for Research in Astronomy, Incorporated, under NASA contract NAS5-26555. 
E. B. J. was supported by grant number HST-GO-15335.002-A from STScI to Princeton University. Based on data obtained from the ESO Science Archive Facility. This work has made use of data from the European Space Agency (ESA) mission {\it Gaia} (\url{https://www.cosmos.esa.int/gaia}), processed by the {\it Gaia} Data Processing and Analysis Consortium (DPAC, \url{https://www.cosmos.esa.int/web/gaia/dpac/consortium}). Funding for the DPAC has been provided by national institutions, in particular the institutions participating in the {\it Gaia} Multilateral Agreement. The background image in Fig. \ref{fig map} is courtesy of NASA/JPL-Caltech/R. Hurt (SSC/Caltech)\\

\textbf{Author contribution} \hspace{0.5cm}  A.D.C. initiated, designed, and directed the project, is the PI of the HST data, analyzed and interpreted the data, developed and applied the main methodology, and wrote the bulk of the manuscript. E.B.J. reduced the HST data and retrieved the UVES data, analyzed and interpreted the data, measured the column densities, developed and applied one of the two methods to measure the metallicity, contributed to the writing, and produced Fig. \ref{fig F* fits}. A.J.F. contributed to the writing and scientific design of the paper. C.L. checked the consistency of the analysis, helped interpreting the data, and contributed to the writing. E.B.J., C.L., A.J.F., and P.P. are co-Is of the HST data. T.R.H. measured the position of our targets within the Galaxy and produced Fig. \ref{fig map} and Extended Data Figure \ref{fig met radial}. C.K. reviewed the depletion methods and assumptions, and collected data from Galactic extinction maps. P.P. contributed to the prioritization of the scientific goals and the writing. J.K.K. assessed the ionization effects and contributed to the writing. J.K.K., T.R.H., C.K., C.L. and A.D.C. measured the column densities towards eight targets with an independent method for a cross-check of the results. All authors participated in the scientific interpretation, edited the manuscript and contributed to its revision.\\

\textbf{Competing Interests} \hspace{0.5cm} The authors declare that they have no competing financial interests.\\

\textbf{Data Availability} \hspace{0.5cm} The observational data used in this work are publicly available in the Mikulski Archive for Space Telescope (HST/STIS data, Program ID 15335, PI De Cia, DOI: 10.17909/t9-r14v-tp03) and the ESO Science Archive Facility (VLT/UVES data \url{http://archive.eso.org/wdb/wdb/adp/phase3_spectral/form?}). The data used in Figures and Tables are available as electronically readable Source Data files, except for the publicly available observational data (Extended Data Figure \ref{fig lines}).

\textbf{Code Availability}  \hspace{0.5cm} The VoigtFit software is publicly available on GitHub  \url{https://github.com/jkrogager/VoigtFit}.

\textbf{Additional Information} \hspace{0.5cm} Correspondence and requests for materials should be addressed to A.D.C.~(email: annalisa.decia@unige.ch). Reprints and permissions information is available at www.nature.com/reprints.\\

\end{flushleft}

 \clearpage    

\section*{Extended Data}

\renewcommand{\figurename}{Extended Data Figure}
\renewcommand{\tablename}{Extended Data Table}

\setcounter{table}{0}
\setcounter{figure}{0}


\begin{figure}
\centering
\includegraphics[width=\textwidth]{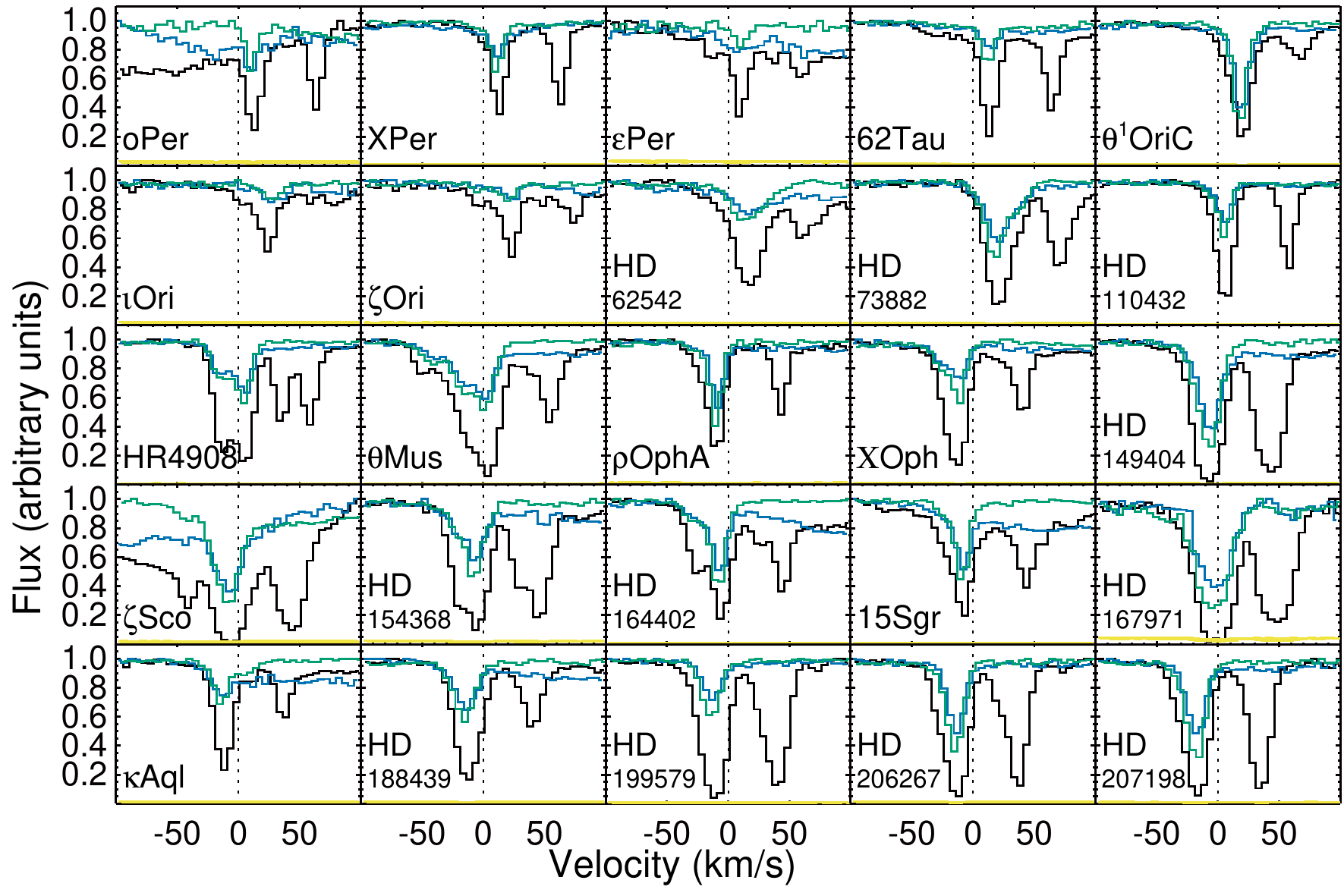}
\caption{\textbf{Line profiles of \znii{} $\lambda$ 2026 (black), \crii{} $\lambda$ 2056 (blue), and \feii{} $\lambda$ 2260 (green) in our sample.} The \mgi{} $\lambda$ 2026 line is separated by $\sim50$~km~s$^{-1}$ from \znii{}. Vertical lines mark the zero-velocity central wavelength of the \znii{} line. The yellow curve shows the $1\,\sigma$ uncertainties.}
\label{fig lines}
\end{figure}

\newpage

\begin{figure}
\centering
\includegraphics[width=\textwidth]{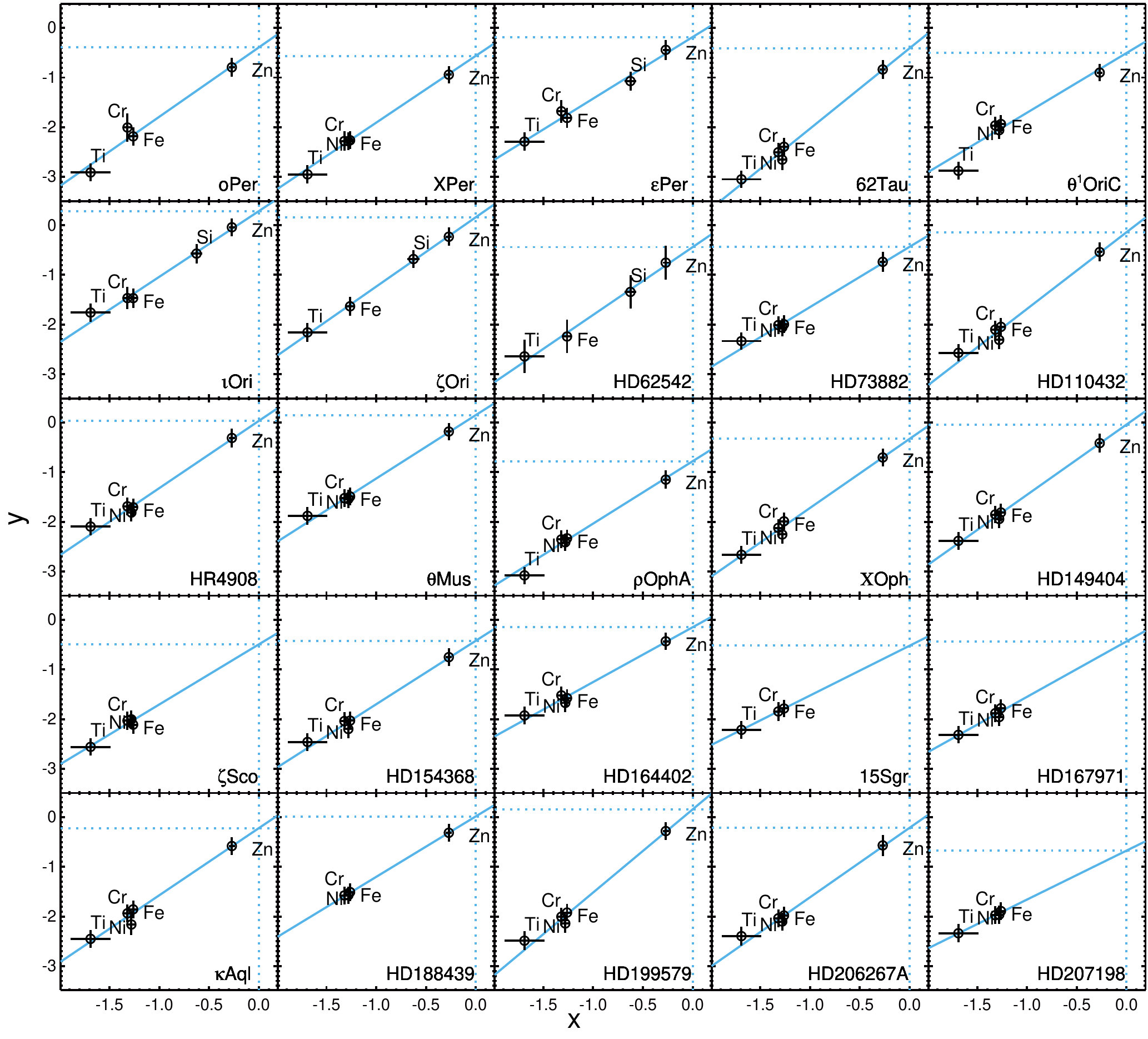}
\caption{\textbf{Determination of the metallicity and strength of depletion with the relative method.} The variables and coefficients of the linear relation are defined in Eqs. \ref{eq a} to \ref{eq y}, where the $y$-intercept gives the [M/H]$_{\rm tot}$ and the slope of the relation the strength of depletion [Zn/Fe]$_{\rm fit}$. The error bars show the $1\,\sigma$ uncertainties.}
\label{fig ZnFe fit}
\end{figure}

\newpage

\begin{figure}
\centering
\includegraphics[width=\textwidth]{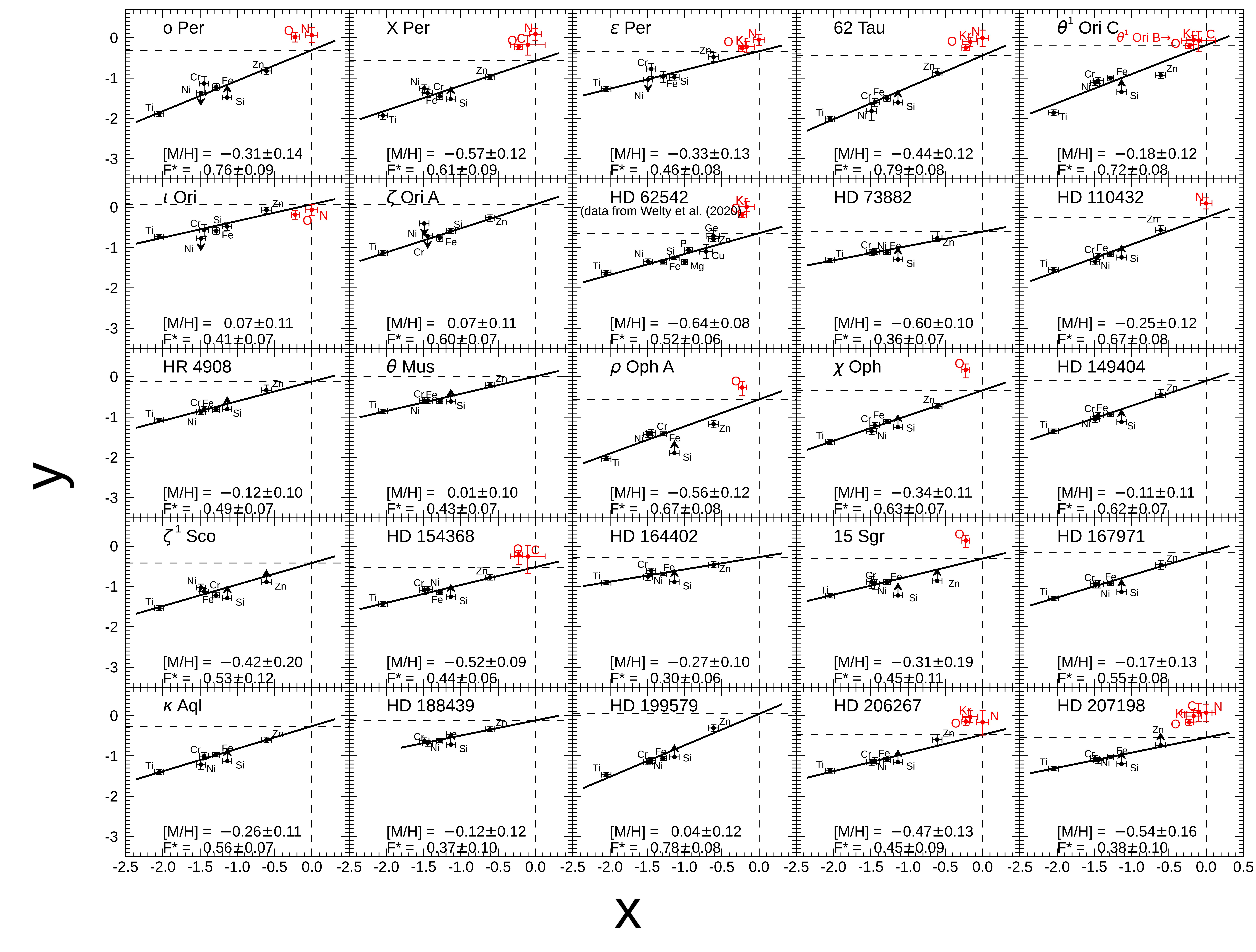}
\caption{\textbf{Determination of [M/H]$_{\rm tot}$ and $F*$ with the $F*$ method.} The variables are described in Eq. \ref{eq F*}. The most volatile elements (red) are taken from the literature (Table \ref{table volatile}) and shown for reference: their discrepancy with respect to the more refractory elements suggests a mix between high-metallicity and pristine gas, see Methods. The error bars show the $1\,\sigma$ uncertainties.}
\label{fig F* fits}
\end{figure}

\clearpage

\begin{figure}
\centering
\includegraphics[width=0.6\textwidth]{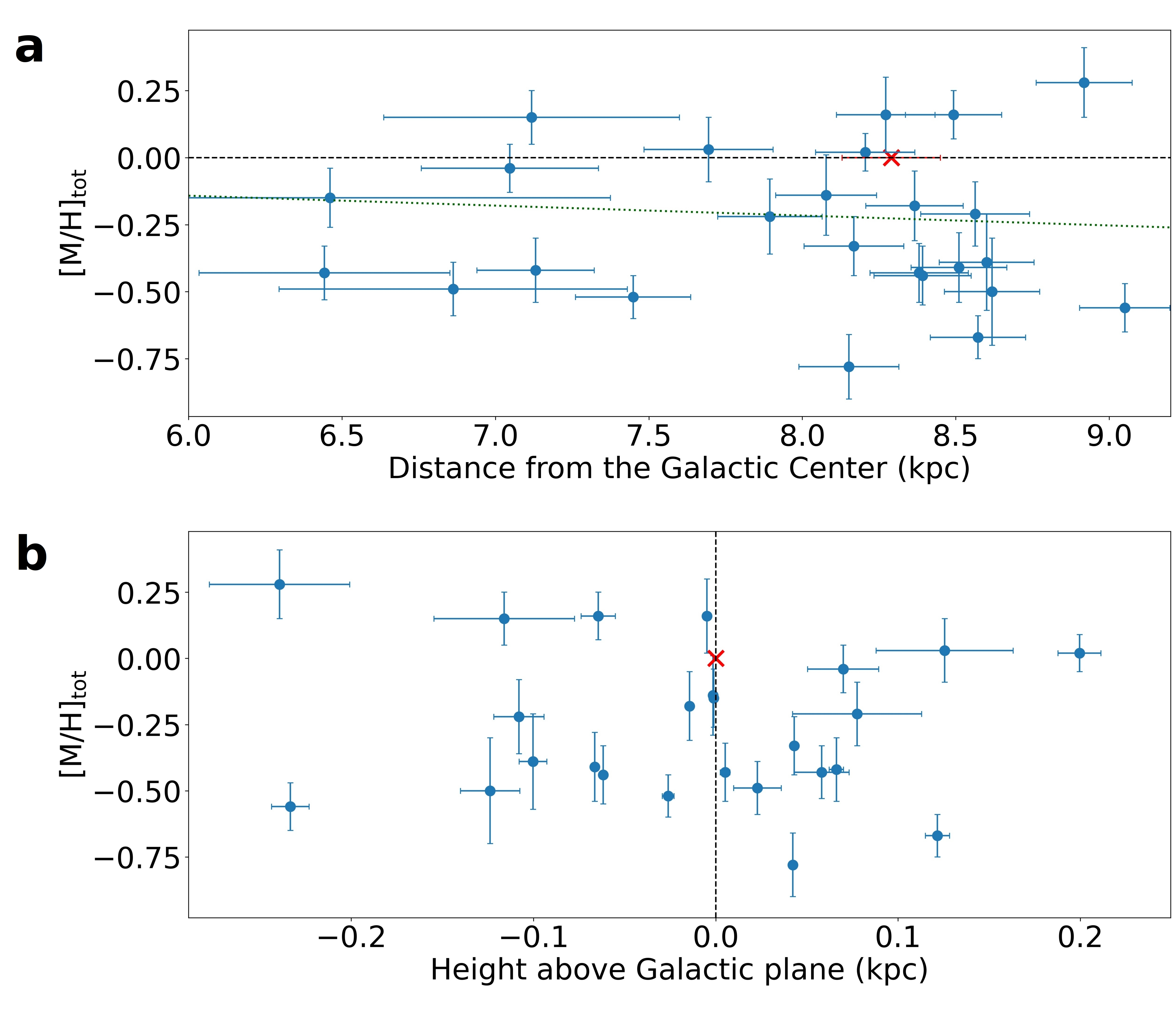}
\caption{\textbf{a) Metallicities towards our targets and their Galactic radii.} The green dotted line shows the metallcitiy gradient measured in \hii{} regions by \cite{Arellano20}, although without dust corrections. The Solar Galactic radius (red cross) is assumed at 8.29~kpc \cite{McMillan11}. The error bars show the $1\,\sigma$ uncertainties. \textbf{b) Metallicities towards our targets and their height above the Galactic disk.} The error bars show the $1\,\sigma$ uncertainties.}
\label{fig met radial}
\end{figure}


\renewcommand{\familydefault}{\sfdefault}
\everymath={\sf}

\clearpage    

\begin{table}
\resizebox{\textwidth}{!}{
\begin{tabular}{@{}l | r r r r r r r r r l@{}}
\hline
               ID &   HD &            Type &  $l$    & $b$      & $r_{\rm Gaia DR2}$ &      $V$ & $E(B-V)$ & $R_V$ &  $\log N$(\hi)&  $\log N$(\hh)    \\
                  &      &                 &  [deg]  & [deg]    & [pc]               &    [mag] & [mag]    & &  &  \\
\hline
         $o$~Per & 23180 &         B1IIIB &$160.364$&$-17.740 $&    $  329\pm   25$ &$   3.83 $&      $0.31\pm0.03$  &    $ 3.11\pm 0.39$  &    $20.82\pm 0.10$  &    $20.60\pm 0.10$ \\
            X~Per & 24534 &         O9.5pe &$163.081$&$-17.136 $&    $  792\pm   35$ &$   6.10 $&      $0.59\pm0.03$  &                 --  &    $20.73\pm 0.10$  &    $20.92\pm 0.10$ \\
   $\epsilon$~Per & 24760 &       B0.5V+A2 &$157.354$&$-10.088 $&    $   82\pm    6$ &$   2.89 $&      $0.10\pm0.03$  &                 --  &    $20.45\pm 0.10$  &    $19.52\pm 0.10$ \\
           62~Tau & 27778 &            B3V &$172.763$&$-17.393 $&    $  222\pm    2$ &$   6.36 $&      $0.37\pm0.03$  &    $ 2.59\pm 0.24$  &    $20.95\pm 0.10$  &    $20.79\pm 0.10$ \\
 $\theta^1$~Ori~C & 37022 &          O6-7p &$209.011$&$-19.384 $&    $  373\pm   49$ &$   5.13 $&      $0.34\pm0.03$  &    $ 5.73\pm 0.69$  &    $21.54\pm 0.10$  &    $17.25\pm 0.10$ \\
      $\iota$~Ori & 37043 &          O9III &$209.522$&$-19.583 $&($  714\pm  115^a$) &$   2.77 $&      $0.07\pm0.03$  &                 --  &    $20.20\pm 0.10$  &    $14.69\pm 0.10$ \\
    $\zeta$~Ori~A & 37742 &        O9.5Ibe &$206.452$&$-16.585 $&($  226\pm   33^a$) &$   2.05 $&      $0.06\pm0.03$  &                 --  &    $20.39\pm 0.10$  &    $15.86\pm 0.10$ \\
         HD~62542 & 62542 &        B5V/B3V &$255.915$&$ -9.237 $&    $  385\pm    4$ &$   8.04 $&      $0.35\pm0.03$  &    $ 2.82\pm 0.24$  &    $20.70\pm 0.18$  &    $20.81\pm 0.20$ \\
         HD~73882 & 73882 &            O8V &$260.182$&$  0.643 $&($  460\pm  234^a$) &$   7.22 $&      $0.70\pm0.03$  &    $ 3.56\pm 0.13$  &    $21.11\pm 0.10$  &    $21.11\pm 0.10$ \\
        HD~110432 &110432 &           B0.5IVpe &$301.958$&$ -0.203 $&    $  416\pm   21$ &$   5.31 $&    ($0.51\pm0.03$)  &                 --  &    $20.85\pm 0.10$  &    $20.64\pm 0.10$ \\
          HR~4908 &112244 &          O9Ibe &$303.553$&$  6.031 $&  ($ 1195\pm  358$) &$   5.32 $&      $0.30\pm0.03$  &                 --  &    $21.08\pm 0.10$  &    $20.14\pm 0.10$ \\
     $\theta$~Mus &113904 &      WC6+O6-7V &$304.675$&$ -2.491 $&  ($ 2671\pm  887$) &$   5.51 $&      $0.25\pm0.03$  &                 --  &    $21.15\pm 0.10$  &    $19.83\pm 0.10$ \\
     $\rho$~Oph~A &147933 &           B2IV &$353.686$&$ 17.687 $&    $  139\pm    3$ &$   5.02 $&      $0.48\pm0.03$  &    $ 5.74\pm 0.40$  &    $21.63\pm 0.10$  &    $20.57\pm 0.10$ \\
       $\chi$~Oph &148184 &         B2IVpe &$357.933$&$ 20.677 $&    $  122\pm    4$ &$   4.42 $&    ($0.52\pm0.03$)  &                 --  &    $21.13\pm 0.10$  &    $20.63\pm 0.10$ \\
        HD~149404 &149404 &          O9Iae &$340.538$&$  3.006 $&  ($ 1333\pm  373$) &$   5.47 $&      $0.68\pm0.03$  &    $ 3.53\pm 0.38$  &    $21.40\pm 0.10$  &    $20.79\pm 0.10$ \\
    $\zeta^1$~Sco &152236 &        B1Ia+pe &$343.028$&$  0.870 $&  ($ 1507\pm  863$) &$   4.73 $&    ($0.68\pm0.03$)  &    $ 3.73\pm 0.39$  &    $21.77\pm 0.10$  &    $20.73\pm 0.10$ \\
        HD~154368 &154368 &           O9Ia &$349.970$&$  3.215 $&    $ 1180\pm   70$ &$   6.13 $&      $0.78\pm0.03$  &    $ 3.33\pm 0.15$  &    $21.00\pm 0.10$  &    $21.16\pm 0.10$ \\
        HD~164402 &164402 &        B0Iab/b &$  7.162$&$ -0.034 $&  ($ 1847\pm 1275$) &$   5.77 $&      $0.28\pm0.03$  &    $ 3.03\pm 0.62$  &    $21.11\pm 0.10$  &    $19.49\pm 0.10$ \\
           15~Sgr &167264 &        O9.7Iab &$ 10.456$&$ -1.741 $&    $  857\pm  106$ &$   5.37 $&      $0.34\pm0.03$  &    $ 3.26\pm 0.31$  &    $21.15\pm 0.10$  &    $20.28\pm 0.10$ \\
        HD~167971 &167971 & O8Iaf(n)+O4/5C &$ 18.251$&$  1.684 $&  ($ 1976\pm  515$) &$   7.45 $&      $1.08\pm0.03$  &    $ 3.44\pm 0.10$  &    $21.60\pm 0.10$  &    $20.85\pm 0.10$ \\
     $\kappa$~Aql &184915 &       B0.5IIIn &$ 31.771$&$-13.287 $&    $  470\pm   60$ &$   4.96 $&      $0.26\pm0.03$  &                 --  &    $20.90\pm 0.10$  &    $20.31\pm 0.10$ \\
        HD~188439 &188439 &      B0.5IIInC &$ 81.772$&$ 10.320 $&    $ 1114\pm   66$ &$   6.28 $&      $0.14\pm0.03$  &                 --  &    $20.78\pm 0.10$  &    $19.95\pm 0.10$ \\
        HD~199579 &199579 &           O6Ve &$ 85.697$&$ -0.300 $&    $  918\pm   50$ &$   5.96 $&      $0.37\pm0.03$  &    $ 3.17\pm 0.69$  &    $21.04\pm 0.10$  &    $20.53\pm 0.10$ \\
        HD~206267 &206267 &O6.0V((f))+O9:V &$ 99.290$&$  3.738 $&  ($ 1190\pm  543$) &$   5.62 $&      $0.53\pm0.03$  &    $ 2.82\pm 0.16$  &    $21.30\pm 0.10$  &    $20.86\pm 0.10$ \\
        HD~207198 &207198 &          O9IIe &$103.136$&$  6.995 $&    $  999\pm   54$ &$   5.95 $&      $0.62\pm0.03$  &    $ 2.77\pm 0.35$  &    $21.34\pm 0.10$  &    $20.83\pm 0.10$ \\

\hline	
\end{tabular}}
\caption{\textbf{Target sample characteristics.} In parenthesis are reported unreliable values of $E(B-V)$, because of contamination by strong H-alpha emission, and unreliable values of the Gaia DR2 distance, because of a large parallax uncertainty (i.e. parallax-to-error $\leq 5$). $^a$ Hipparcos distance. \hi{} and \hh{} values are taken from \cite{Welty10}, with the exception of HD~62542 \cite{Welty20}. $E(B-V)$ values are taken from \cite{Diplas94} and $R_V$ from \cite{Valencic04}.}
\label{table sample}
\end{table}

 \clearpage

\begin{table}
\resizebox{\textwidth}{!}{
\begin{tabular}{@{}l | r r r r r r r r r@{}}
\hline
          & \multicolumn{9}{c}{$\log N$} \\
    ID         &   \mgi     &  \alii    &  \siii  &  \tiii    &  \crii         &  \feii   & \coii    &  \niii        &  \znii       \\
\hline    
           $o$~Per  & $ 13.67\pm  0.13$  & $        > 12.81$  & $        > 15.00$  & $ 11.16\pm  0.05$  & $ 12.93\pm  0.22$  & $ 14.45\pm  0.06$  & $        < 13.05$  & $        < 13.14$    & $ 13.00\pm  0.07$   \\
             X~Per  & $ 13.68\pm  0.05$  & $        > 13.09$  & $        > 15.14$& $ 11.30\pm  0.07^a$  & $ 12.84\pm  0.09$  & $ 14.55\pm  0.03$  & $        < 12.13$  & $ 13.36\pm  0.05$    & $ 13.03\pm  0.04$   \\
    $\epsilon$~Per  & $             --$  & $        > 12.69$  & $ 14.88\pm  0.06$& $ 11.16\pm  0.07^a$  & $ 12.64\pm  0.14$  & $ 14.20\pm  0.09$  & $        < 12.89$  & $        < 12.85$    & $ 12.73\pm  0.10$   \\
            62~Tau  & $ 13.40\pm  0.06$  & $        > 12.57$  & $        > 15.04$  & $ 11.18\pm  0.03$  & $ 12.60\pm  0.06$  & $ 14.40\pm  0.04$  & $        < 12.65$  & $ 12.97\pm  0.11$    & $ 13.12\pm  0.08$   \\
  $\theta^1$~Ori~C  & $ 13.86\pm  0.08$  & $        > 13.24$  & $        > 15.52$  & $ 11.57\pm  0.05$  & $ 13.35\pm  0.02$  & $ 15.07\pm  0.02$  & $        < 12.59$  & $ 13.77\pm  0.04$    & $ 13.27\pm  0.04$   \\
       $\iota$~Ori  & $             --$  & $        > 13.09$  & $ 15.04\pm  0.09$& $ 11.35\pm  0.07^a$  & $ 12.51\pm  0.14$  & $ 14.20\pm  0.09$  & $        < 12.71$  & $        < 12.77$    & $ 12.79\pm  0.04$   \\
     $\zeta$~Ori~A  & $        < 13.44$  & $        > 13.10$  & $ 15.12\pm  0.05$& $ 11.14\pm  0.07^a$  & $        < 12.55$  & $ 14.22\pm  0.06$  & $             --$  & $        < 13.34$    & $ 12.79\pm  0.07$   \\
          HD~62542  & $ 13.81\pm  0.04$  & $             --$  & $ 15.32\pm  0.02$& $ 11.52\pm  0.03^a$  & $             --$  & $ 14.48\pm  0.02$  & $        < 12.60$  & $ 13.26\pm  0.04$    & $ 13.13\pm  0.05$   \\
          HD~73882  & $ 13.41\pm  0.14$  & $        > 13.39$  & $        > 15.61$  & $ 12.16\pm  0.02$  & $ 13.35\pm  0.02$  & $ 15.06\pm  0.02$  & $ 12.83\pm  0.22$  & $ 13.82\pm  0.04$    & $ 13.48\pm  0.10$   \\
         HD~110432  & $ 13.62\pm  0.03$  & $        > 12.97$  & $        > 15.27$  & $ 11.53\pm  0.03$  & $ 12.87\pm  0.03$  & $ 14.62\pm  0.03$  & $        < 12.25$  & $ 13.19\pm  0.04$    & $ 13.29\pm  0.08$   \\
           HR~4908  & $ 13.73\pm  0.06$  & $        > 13.30$  & $        > 15.68$  & $ 11.98\pm  0.02$  & $ 13.26\pm  0.02$  & $ 14.93\pm  0.02$  & $        < 12.40$  & $ 13.65\pm  0.04$    & $ 13.49\pm  0.08$   \\
      $\theta$~Mus  & $ 13.39\pm  0.14$  & $        > 13.58$  & $        > 15.89$  & $ 12.22\pm  0.02$  & $ 13.44\pm  0.03$  & $ 15.17\pm  0.02$  & $ 13.12\pm  0.13$  & $ 13.94\pm  0.06$    & $ 13.64\pm  0.03$   \\
      $\rho$~Oph~A  & $ 13.54\pm  0.04$  & $        > 12.98$  & $        > 15.12$  & $ 11.53\pm  0.02$  & $ 13.12\pm  0.02$  & $ 14.84\pm  0.02$  & $        < 12.60$  & $ 13.58\pm  0.03$    & $ 13.18\pm  0.07$   \\
        $\chi$~Oph  & $ 13.49\pm  0.06$  & $        > 13.05$  & $        > 15.41$  & $ 11.59\pm  0.03$  & $ 12.99\pm  0.03$  & $ 14.82\pm  0.02$  & $        < 12.73$  & $ 13.38\pm  0.04$    & $ 13.27\pm  0.03$   \\
         HD~149404  & $ 14.16\pm  0.03$  & $        > 13.40$  & $        > 15.77$  & $ 12.10\pm  0.02$  & $ 13.49\pm  0.02$  & $ 15.23\pm  0.02$  & $ 13.14\pm  0.14$  & $ 13.93\pm  0.03$    & $ 13.79\pm  0.09$   \\
     $\zeta^1$~Sco  & $        > 14.29$  & $        > 13.58$  & $        > 15.87$  & $ 12.19\pm  0.03$  & $ 13.59\pm  0.05$  & $ 15.20\pm  0.05$  & $             --$  & $ 14.13\pm  0.06$    & $        > 13.61$   \\
         HD~154368  & $ 14.04\pm  0.09$  & $        > 13.28$  & $        > 15.65$  & $ 12.04\pm  0.04$  & $ 13.32\pm  0.05$  & $ 15.03\pm  0.03$  & $        < 12.67$  & $ 13.69\pm  0.05$    & $ 13.47\pm  0.04$   \\
         HD~164402  & $ 13.40\pm  0.18$  & $        > 13.38$  & $        > 15.56$  & $ 12.11\pm  0.02$  & $ 13.38\pm  0.02$  & $ 15.02\pm  0.02$  & $        < 13.10$  & $ 13.75\pm  0.05$    & $ 13.33\pm  0.04$   \\
            15~Sgr  & $ 13.42\pm  0.14$  & $        > 13.49$  & $        > 15.35$  & $ 11.95\pm  0.03$  & $ 13.18\pm  0.08$  & $ 14.94\pm  0.03$  & $             --$  & $ 13.72\pm  0.14$    & $        > 13.05$   \\
         HD~167971  & $ 14.25\pm  0.05$  & $        > 13.45$  & $        > 15.92$  & $ 12.32\pm  0.02$  & $ 13.62\pm  0.03$  & $ 15.42\pm  0.03$  & $        < 13.35$  & $ 14.07\pm  0.04$    & $        > 13.70$   \\
      $\kappa$~Aql  & $ 13.32\pm  0.13$  & $        > 13.23$  & $        > 15.27$  & $ 11.54\pm  0.03$  & $ 12.92\pm  0.05$  & $ 14.69\pm  0.03$  & $ 13.01\pm  0.23$  & $ 13.22\pm  0.11$    & $ 13.13\pm  0.04$   \\
         HD~188439  & $ 13.28\pm  0.09$  & $        > 13.22$  & $        > 15.49$  & $             --$  & $ 13.09\pm  0.03$  & $ 14.85\pm  0.02$  & $        < 12.54$  & $ 13.62\pm  0.03$    & $ 13.21\pm  0.02$   \\
         HD~199579  & $ 13.93\pm  0.06$  & $        > 13.18$  & $        > 15.54$& $ 11.67\pm  0.07^a$  & $ 13.02\pm  0.04$  & $ 14.80\pm  0.02$  & $        < 12.24$  & $ 13.40\pm  0.06$    & $ 13.60\pm  0.05$   \\
         HD~206267  & $ 14.06\pm  0.03$  & $        > 13.23$  & $        > 15.70$& $ 12.05\pm  0.07^a$  & $ 13.27\pm  0.02$  & $ 15.02\pm  0.02$  & $        < 12.73$  & $ 13.73\pm  0.03$    & $ 13.60\pm  0.12$   \\
         HD~207198  & $ 14.21\pm  0.06$  & $        > 13.36$  & $        > 15.67$& $ 12.12\pm  0.07^a$  & $ 13.35\pm  0.03$  & $ 15.13\pm  0.02$  & $ 13.19\pm  0.13$  & $ 13.88\pm  0.05$    & $        > 13.47$   \\     
\hline
\end{tabular}}
\caption{\textbf{Column densities.} They are measured from the HST/STIS data of the 25 stars in our sample. $^a$Measurements from \cite{Welty10}.}
\label{table cold}
\end{table}

 \clearpage

\begin{table}
\centering
\resizebox{\textwidth}{!}{
\begin{tabular}{@{}l | r r r r r r r r@{}}
\hline
                ID & $\log N(\mbox{H})$& [Zn/H]  & [Zn/Fe]          & [Zn/Fe]$_{\rm fit}$& [$M$/H]$_{\rm tot, [X/Y]}$ & $F*$   &  [$M$/H]$_{{\rm tot}, F*}$ & Z-test\\   
                \hline
          $o$~Per  & $21.16\pm 0.17$  & $-0.79\pm 0.16$  & $ 1.40\pm 0.09$  & $ 1.40\pm 0.11$  & $-0.39\pm 0.18$  &  $ 0.76\pm0.09$  &  $-0.31\pm0.14$  &-0.36 \\
            X~Per  & $21.34\pm 0.17$  & $-0.94\pm 0.15$  & $ 1.33\pm 0.05$  & $ 1.35\pm 0.07$  & $-0.56\pm 0.09$  &  $ 0.61\pm0.09$  &  $-0.57\pm0.12$  & 0.05 \\
   $\epsilon$~Per  & $20.54\pm 0.17$  & $-0.44\pm 0.17$  & $ 1.38\pm 0.14$  & $ 1.24\pm 0.13$  & $-0.18\pm 0.13$  &  $ 0.46\pm0.08$  &  $-0.33\pm0.13$  & 0.81 \\
           62~Tau  & $21.33\pm 0.17$  & $-0.84\pm 0.16$  & $ 1.56\pm 0.09$  & $ 1.62\pm 0.10$  & $-0.41\pm 0.13$  &  $ 0.79\pm0.08$  &  $-0.44\pm0.12$  & 0.18 \\
 $\theta^1$~Ori~C  & $21.54\pm 0.17$  & $-0.90\pm 0.15$  & $ 1.04\pm 0.05$  & $ 1.20\pm 0.06$  & $-0.50\pm 0.20$  &  $ 0.72\pm0.08$  &  $-0.18\pm0.12$  &-1.41 \\
      $\iota$~Ori  & $20.20\pm 0.17$  & $-0.04\pm 0.15$  & $ 1.43\pm 0.10$  & $ 1.32\pm 0.10$  & $ 0.28\pm 0.13$  &  $ 0.41\pm0.07$  &  $ 0.07\pm0.11$  & 1.27 \\
    $\zeta$~Ori~A  & $20.39\pm 0.17$  & $-0.23\pm 0.16$  & $ 1.41\pm 0.10$  & $ 1.40\pm 0.11$  & $ 0.16\pm 0.09$  &  $ 0.61\pm0.07$  &  $ 0.11\pm0.11$  & 0.37 \\
         HD~62542  & $21.25\pm 0.34$  & $-0.75\pm 0.15$  & $ 1.50\pm 0.05$  & $ 1.37\pm 0.09$  & $-0.44\pm 0.11$  &  $ 0.48\pm0.06$  &  $-0.69\pm0.07$  & 1.90 \\
         HD~73882  & $21.59\pm 0.17$  & $-0.74\pm 0.17$  & $ 1.26\pm 0.10$  & $ 1.21\pm 0.11$  & $-0.43\pm 0.11$  &  $ 0.36\pm0.07$  &  $-0.60\pm0.10$  & 1.14 \\
        HD~110432  & $21.20\pm 0.17$  & $-0.54\pm 0.16$  & $ 1.51\pm 0.08$  & $ 1.55\pm 0.10$  & $-0.14\pm 0.15$  &  $ 0.67\pm0.08$  &  $-0.25\pm0.08$  & 0.65 \\
          HR~4908  & $21.17\pm 0.17$  & $-0.31\pm 0.16$  & $ 1.40\pm 0.08$  & $ 1.36\pm 0.10$  & $ 0.03\pm 0.12$  &  $ 0.49\pm0.07$  &  $-0.12\pm0.10$  & 0.98 \\
     $\theta$~Mus  & $21.19\pm 0.17$  & $-0.18\pm 0.14$  & $ 1.32\pm 0.04$  & $ 1.28\pm 0.06$  & $ 0.15\pm 0.10$  &  $ 0.43\pm0.07$  &  $ 0.01\pm0.10$  & 1.01 \\
     $\rho$~Oph~A  & $21.70\pm 0.17$  & $-1.15\pm 0.16$  & $ 1.19\pm 0.07$  & $ 1.26\pm 0.08$  & $-0.78\pm 0.12$  &  $ 0.67\pm0.08$  &  $-0.56\pm0.12$  &-1.30 \\
       $\chi$~Oph  & $21.34\pm 0.17$  & $-0.70\pm 0.14$  & $ 1.29\pm 0.04$  & $ 1.39\pm 0.07$  & $-0.33\pm 0.11$  &  $ 0.63\pm0.07$  &  $-0.34\pm0.11$  & 0.08 \\
        HD~149404  & $21.57\pm 0.17$  & $-0.41\pm 0.17$  & $ 1.41\pm 0.09$  & $ 1.42\pm 0.10$  & $-0.04\pm 0.09$  &  $ 0.62\pm0.07$  &  $-0.11\pm0.11$  & 0.51 \\
    $\zeta^1$~Sco  & $21.84\pm 0.17$  &       $ >-0.86$  &       $ > 1.26$  & $ 1.21\pm 0.88$  & $-0.49\pm 0.10$  &  $ 0.53\pm0.12$  &  $-0.42\pm0.20$  &-0.33 \\
        HD~154368  & $21.59\pm 0.17$  & $-0.75\pm 0.15$  & $ 1.29\pm 0.04$  & $ 1.28\pm 0.07$  & $-0.42\pm 0.12$  &  $ 0.44\pm0.06$  &  $-0.52\pm0.09$  & 0.67 \\
        HD~164402  & $21.13\pm 0.17$  & $-0.43\pm 0.15$  & $ 1.15\pm 0.04$  & $ 1.11\pm 0.06$  & $-0.15\pm 0.11$  &  $ 0.30\pm0.06$  &  $-0.27\pm0.10$  & 0.85 \\
           15~Sgr  & $21.25\pm 0.17$  &       $ >-0.83$  &       $ > 0.96$  & $ 1.00\pm 0.51$  & $-0.52\pm 0.08$  &  $ 0.45\pm0.11$  &  $-0.31\pm0.19$  &-1.01 \\
        HD~167971  & $21.73\pm 0.17$  &       $ >-0.66$  &       $ > 1.12$  & $ 1.12\pm 1.37$  & $-0.43\pm 0.10$  &  $ 0.55\pm0.08$  &  $-0.17\pm0.13$  &-1.60 \\
     $\kappa$~Aql  & $21.08\pm 0.17$  & $-0.58\pm 0.15$  & $ 1.28\pm 0.05$  & $ 1.35\pm 0.07$  & $-0.22\pm 0.14$  &  $ 0.56\pm0.07$  &  $-0.26\pm0.11$  & 0.23 \\
        HD~188439  & $20.89\pm 0.17$  & $-0.31\pm 0.14$  & $ 1.20\pm 0.03$  & $ 1.22\pm 0.06$  & $ 0.02\pm 0.07$  &  $ 0.37\pm0.10$  &  $-0.12\pm0.12$  & 0.97 \\
        HD~199579  & $21.25\pm 0.17$  & $-0.28\pm 0.15$  & $ 1.65\pm 0.06$  & $ 1.68\pm 0.08$  & $ 0.16\pm 0.14$  &  $ 0.78\pm0.08$  &  $ 0.04\pm0.12$  & 0.65 \\
        HD~206267  & $21.54\pm 0.17$  & $-0.57\pm 0.19$  & $ 1.42\pm 0.12$  & $ 1.40\pm 0.13$  & $-0.21\pm 0.12$  &  $ 0.45\pm0.09$  &  $-0.47\pm0.13$  & 1.45 \\
        HD~207198  & $21.55\pm 0.17$  &       $ >-0.71$  &       $ > 1.19$  & $ 0.99\pm 0.58$  & $-0.67\pm 0.08$  &  $ 0.38\pm0.10$  &  $-0.54\pm0.16$  &-0.74 \\
        \hline
\end{tabular}}
\caption{\textbf{Metallicities of the neutral ISM.} These are derived using dust corrections from the relative method ([$M$/H]$_{\rm tot, [X/Y]}$) and the $F*$ method [$M$/H]$_{{\rm tot}, F*}$). The last column reports the difference between the two in $\sigma$ levels from a Z-test.}
\label{table met results}
\end{table}

\clearpage

\begin{table}
\centering
\begin{tabular}{l | c c r}
\hline
Ion & $\lambda$ & $\log \lambda f$  & Ref.\\
    &    [\AA{}]&          \\
\hline
\mgi{}  & 1827.935  & 1.677  & \cite{Cashman17} \\ 
\mgi{}  & 2026.477	 & 2.36   & \cite{Cashman17} \\ 
\alii{} & 1670.787	 & 3.463  & \cite{Cashman17} \\ 
\siii{} & 1808.013	 & 0.646  & \cite{Cashman17} \\ 
\tiii{} & 3384.740	 & 3.127  & \cite{Cashman17} \\ 
\tiii{} & 3242.929	 & 2.919  & \cite{Cashman17} \\ 
\tiii{} & 3230.131	 & 2.377  & \cite{Cashman17} \\ 
\crii{} & 2056.254	 & 2.351  & \cite{Cashman17} \\ 
\crii{} & 2062.234	 & 2.224  & \cite{Cashman17} \\ 
\crii{} & 2066.161	 & 2.015  & \cite{Cashman17} \\ 
\feii{} & 2344.214	 & 2.427  & \cite{Cashman17,Morton03} \\ 
\feii{} & 2260.780	 & 0.742  & \cite{Cashman17,Morton03} \\ 
\feii{} & 2249.877	 & 0.612  & \cite{Cashman17,Morton03} \\ 
\coii{} & 2012.161	 & 1.87   & \cite{Cashman17} \\ 
\coii{} & 1941.280	 & 1.82   & \cite{Cashman17} \\ 
\niii{} & 1741.549	 & 1.876  & \cite{Boisse19} \\ 
\niii{} & 1709.600	 & 1.735  & \cite{Boisse19} \\ 
\niii{} & 1751.910	 & 1.691  & \cite{Boisse19} \\ 
\znii{} & 2026.136	 & 3.106  & \cite{Kisielius15} \\ 
\znii{} & 2062.664	 & 2.804  & \cite{Kisielius15} \\ 
\hline	
\end{tabular}
\caption{\textbf{Absorption lines that we use in this work and their oscillator strengths}. For Ni transitions, we used the $f$-values from \cite{Boisse19} based on observational data, which agree with \cite{Jenkins06}, instead of the theoretical measurements of \cite{Cashman17}.}
\label{table lines}
\end{table}

\clearpage

\begin{table}
\centering
\begin{tabular}{l | r r}
\hline
$X$  & $A2_X$ & $B2_X$ \\
\hline
 Mg& $-0.07\pm 0.05$ & $-0.61\pm 0.05 $ \\
 Al& $-0.27\pm 0.00$ & $ 0.00\pm 0.00 $ \\
 Si& $-0.10\pm 0.03$ & $-0.63\pm 0.06 $ \\
 Cr& $ 0.13\pm 0.03$ & $-1.32\pm 0.04 $ \\
 Fe& $-0.01\pm 0.03$ & $-1.26\pm 0.04 $ \\
 Ni& $ 0.09\pm 0.03$ & $-1.28\pm 0.04 $ \\
 Zn& $ 0.00\pm 0.01$ & $-0.27\pm 0.03 $ \\
 Ti& $-0.04\pm 0.01$ & $-1.69\pm 0.20 $ \\
\hline	
\end{tabular}
\caption{\textbf{Coefficients used in Eqs. \ref{eq x} and \ref{eq y}.} They are taken from the literature, from \cite{DeCia16} except for Ni and Ti (from \cite{Wiseman17}), but here updated for the most recent $f$-values, listed in Extended Data Table \ref{table lines}.}
\label{table AB}
\end{table}

\clearpage

\begin{table}
\resizebox{\textwidth}{!}{
\begin{tabular}{l | r r r  r r r  r r r  r r r }
\hline
          & \multicolumn{12}{c}{$\log N$} \\
    ID         &  \oi$_{\rm MIN}$ & \oi & \oi$_{\rm MAX}$  &  \kri$_{\rm MIN}$ & \kri & \kri$_{\rm MAX}$  &  \cii$_{\rm MIN}$ & \cii & \cii$_{\rm MAX}$   &  \nii$_{\rm MIN}$ & \nii & \nii$_{\rm MAX}$     \\        
\hline  
             $o$~Per & 17.82 & 17.93 & 18.02 &    -- &    -- &    -- &    -- &    -- &    -- & 16.87 & 17.02 & 17.17 \\
               X~Per & 17.85 & 17.87 & 17.89 &    -- &    -- &    -- & 17.35 & 17.51 & 17.62 & 17.14 & 17.22 & 17.30 \\
      $\epsilon$~Per & 16.98 & 17.03 & 17.07 & 11.38 & 11.46 & 11.53 &    -- &    -- &    -- & 16.22 & 16.28 & 16.35 \\
              62~Tau & 17.79 & 17.83 & 17.87 & 12.31 & 12.37 & 12.41 &    -- &    -- &    -- & 16.95 & 17.11 & 17.27 \\
    $\theta^1$~Ori~C & 18.06 & 18.09 & 18.12 & 12.59 & 12.63 & 12.67 & 17.64 & 17.82 & 17.94 &    -- &    -- &    -- \\
         $\iota$~Ori & 16.67 & 16.76 & 16.84 &    -- &    -- &    -- &    -- &    -- &    -- & 15.86 & 15.93 & 15.99 \\
            HD~62542 & 17.80 & 17.82 & 17.80 & 12.30 & 12.41 & 12.52 &    -- &    -- &    -- &    -- &    -- &    -- \\
           HD~110432 &    -- &    -- &    -- &    -- &    -- &    -- &    -- &    -- &    -- & 17.02 & 17.09 & 17.15 \\
        $\rho$~Oph~A & 17.98 & 18.18 & 18.31 &    -- &    -- &    -- &    -- &    -- &    -- &    -- &    -- &    -- \\
          $\chi$~Oph & 18.07 & 18.26 & 18.39 &    -- &    -- &    -- &    -- &    -- &    -- &    -- &    -- &    -- \\
           HD~154368 & 17.88 & 18.11 & 18.21 &    -- &    -- &    -- & 17.30 & 17.68 & 17.88 &    -- &    -- &    -- \\
              15~Sgr & 17.98 & 18.14 & 18.26 &    -- &    -- &    -- &    -- &    -- &    -- &    -- &    -- &    -- \\
           HD~206267 & 18.07 & 18.10 & 18.13 & 12.49 & 12.59 & 12.67 &    -- &    -- &    -- & 16.87 & 17.16 & 17.43 \\
           HD~207198 & 18.12 & 18.15 & 18.18 & 12.58 & 12.67 & 12.74 & 17.84 & 17.98 & 18.09 & 17.22 & 17.41 & 17.59 \\
\hline	
\end{tabular}}
\caption{\textbf{Column densities of the volatile elements.} These values are taken from \cite{Jenkins09} and \cite{Jenkins19}.}
\label{table volatile}
\end{table}

\clearpage

\newpage

\section*{Addendum}

In this Article, we stated that significant amounts of low-metallicity gas in the neutral interstellar medium (ISM) are needed to explain the observed abundance patterns. We clarify here that the exact amount of low-metallicity gas present along each line of sight is difficult to quantify because several assumptions are required to derive the gas mass from the column densities measured along the line of sight. The volatile elements often deviate from the linear fit to the more refractory elements in the abundance patterns (Extended Data Fig. 3 of the original Article). As suggested in the Article, this indicates that there must be a mixture of different gas types along many lines of sight. Near-solar metallicity gas could give rise to the volatile elements, while the low-metallicity nearly dust-free gas could dominate the abundance patterns of the refractory elements. The exact composition of this gas mixture is hard to determine. Our results favour the possibility that an amount between a few per cent and almost half of the gas has a low metallicity, but we cannot rule out a mixture of solar-metallicity gas having vastly different levels of depletion onto dust grains. These results do not contradict the observations that HII regions and OB stars show smaller scatter in metallicity, particularly if the mass contribution of the low-metallicity gas is small, and in general given that some of the neutral gas is in an extended phase of the ISM rather than tracing denser (and possibly more mixed) star-forming environments. Our results do show, however, that low-metallicity gas is present in the ISM, the mass of which is unconstrained so far. Our observations highlight the variety in chemical enrichment of the neutral ISM, both in terms of metallicity and dust depletion. A more complete analysis of the chemical properties of the ISM is under way

\end{document}